\documentclass[12pt,preprint]{aastex}
\begin{document}

\title{Beryllium, Oxygen and Iron Abundances in Extremely Metal-Deficient
Stars} 
\author{Jeffrey A. Rich\altaffilmark{1} \& Ann Merchant Boesgaard\altaffilmark{1} }
\affil{Institute for Astronomy, University of Hawai`i at M\-anoa, \\ 2680
Woodlawn Drive, Honolulu, HI {\ \ }96822 \\ } \email{jrich@ifa.hawaii.edu}
\email{boes@ifa.hawaii.edu}

\altaffiltext{1}{Visiting Astronomer, W.~M.~Keck Observatory jointly operated
 by the California Institute of Technology and the University of California.}

\begin{abstract}
The abundance of beryllium in the oldest, most metal-poor stars acts as a
probe of early star formation and Galactic chemical evolution.  We have
analyzed high-resolution, high signal-to-noise Keck/HIRES spectra of 24 stars
with [Fe/H] from $-$2.3 to $-$3.5 in order to determine the history of Be
abundance and explore the possibility of a Be plateau.  We have determined
stellar parameters of our sample spectroscopically, using equivalent widths of
Fe I, Ti I and Ti II lines.  We have determined O abundances from three OH
features which occur in the same spectral region; this region is relatively
uncrowded and has a well-determined continuum in these very/extremely
metal-poor stars.  We have supplemented this sample with reanalyzed spectra of
25 stars from previous observations so that our total sample ranges in [Fe/H]
from $-$0.5 to $-$3.5.  Our results indicate that the relationship between Be
and [Fe/H] continues to lower metallicities with a slope of 0.92 $\pm$0.04.
Although there is no indication of a plateau with constant Be abundance, the
four lowest metallicity stars (below [Fe/H] of $-$3.0) do show a Be
enhancement relative to Fe at the 1$\sigma$ level.  A single relationship
between Be and [O/H] has a slope of 1.21 $\pm$0.08, but there is also a good
fit with two slopes: 1.59 above [O/H] = $-$1.6 and 0.74 for stars with [O/H]
below $-$1.6. This change in slope could result from a change in the dominant
production mechanism for Be. In the era of the formation of the more metal
poor stars Be would be formed by acceleration of CNO atoms in the vicinity of
SNII and in later times by high-energy cosmic rays bombarding CNO in the
ambient interstellar gas.  We find an excellent correlation between [Fe/H] and
[O/H] and show that [O/Fe] is near +1.0 at [Fe/H] = $-$3.5 declining to 0 at
[Fe/H] = 0.
\end{abstract} 

\keywords{stars: abundances; stars: evolution; stars: late-type; stars
Population II; Galaxy: halo; Galaxy: disk}

\section{Introduction}

The abundances of the rare light elements, Li, Be, and B, have revealed an
amazing amount of information about an array of astrophysical issues.  These
include cosmology and nucleosynthesis in the Big Bang, Galactic chemical
evolution, cosmic ray theory, and stellar interiors and stellar evolution.  It
is important to add more observations of Be to the large number of Li
observations in order to make new progress in these research fields.
Beryllium is the least complicated of the three light elements because it has
only one known source of production and only one stable isotope, $^9$Be.  It
is also less susceptible to depletion than Li in stellar interiors.

Observations of Primas et al.~(2000a, 2000b) of Be in very metal poor stars
led them to suggest the possibility of a Be plateau at very low [Fe/H],
possibly similar to the Spite lithium plateau (Spite \& Spite 1982).  The Li
plateau in older stars is attributed to nucleosynthesis of Li during the Big
Bang.  If such a plateau exists, could it be similar to the Li plateau?
Whereas the Li plateau is produced by nuclear reactions during the Big Bang,
very little Be is produced in standard Big Bang nucleosynthesis (BBN):
N(Be)/N(H) = 10$^{-17}$ (Thomas et al.~1994).  Models that include
inhomogeneities in the early universe could produce a different mixture of
light elements from the standard models (e.g. Malaney \& Mathews 1992).
Inhomogeneous models can create far more Be than standard models and the
predictions of Orito et al.~(1997) are 10$^{-15}$ to 10$^{-14}$; the implied
plateau corresponds to 5 x 10$^{-14}$, which is tantalizingly close to the
observed values.

Another possible cause of a Be ``plateau'' could be the creation of Be through
spallation reactions in contained superbubbles created by multiple supernovae
(Parizot 2000) or near hypernovae (Fields et al.~2002, Nakamura et al.~2006).
Enrichment of the ISM in Be by such processes early in the history of the
galaxy could result in a detectable plateau.

The lowest metal stars ([Fe/H] $<$ $-$3.0) were formed early in the evolution
of the Galaxy.  By tracing the abundance of Be in these old stars,
insight can be gained into the chemical history of the Galaxy.  Beryllium is
formed mainly through spallation reactions, which was proposed by Reeves,
Fowler \& Hoyle (1970) and further described by Meneguzzi, Audouze \& Reeves
(1971).  The basic process is that high energy ($\sim$150 MeV) protons and
neutrons bombard interstellar nuclei of C, N and O creating lighter isotopes.
It has also been suggested that the ``bullets'' and the ``targets'' might be
reversed near supernovae where C, N, and O nuclei could be accelerated into
the local interstellar gas including protons and neutrons (see, for example,
Duncan et al.~1997, 1998, Lemoine, Vangioni-Flam \& Cass\'e 1998). 

Recent studies of Be in low metal stars include Boesgaard, Levesque \& Rich
(2008), Tan, Shi \& Zhao (2009) and Smiljanic et al.~(2009).  All of these
studies examine the Galactic evolution of Be through the relationships between
the abundances of Be with Fe and Be with O and/or alpha-elements.  The large
study of Smiljanic et al.~(2009) includes 90 stars and the determination of Be
abundances of 73 halo and thick disk stars.  They relied on Hipparcos
parallaxes to find log $g$, and adopted published values for temperature,
[Fe/H], [$\alpha$/Fe], and microturbulent velocity.  Their stars range in
[Fe/H] from $-$0.5 to $-$2.5.

In this work we have obtained high signal-to-noise (S/N), high resolution
spectra with the upgraded HIRES on the Keck I telescope of 24 stars covering
values of [Fe/H] of $-$2.32 to $-$3.45 to examine Be abundances at as low
metallicities as possible.  We supplement this sample with reanalyzed
Keck/HIRES data from Boesgaard et al.~(1999a) and Boesgaard (2007) and
Subaru/HDs data from Boesgaard \& Novicki (2006).  Our total sample is 49
dwarfs and subgiants ranging in [Fe/H] from $-$0.5 to $-$3.5, with 34 stars
below $-$2.0.  This is the most metal-poor set of stars which have been
analyzed for Be abundances.  Another difference from previous studies is that
we have determined parameters spectroscopically.

\section{Observations and Data Reduction}
The Be II resonance lines, located in the UV spectral region at 3130.421 and
3131.065 \AA, were used to determine the Be abundances.  High S/N and
high-resolution spectra are required to observe the Be doublet, which is weak
in low metallicity stars.  The upgraded High Resolution Echelle Spectrograph
(HIRES) instrument on the Keck I telescope achieves the necessary requirements
(Vogt et al.~1994).  The spectral resolution for our setup was 48,000, and the
light collecting area of the 10 m Keck telescope enables us to obtain S/N near
100 per pixel for our metal-poor sample.  The pixel size on the upgraded CCD
is 15 $\mu$.  Such stars are uncommon and thus most are faint, especially in
the UV wavelength region of the Be II lines.  The observations reported here
were carried out over 11 nights using the blue cross-disperser.  The spectral
range is approximately 3000 to 6000 \AA.  The new HIRES CCD is composed of
three chips, one each for UV/blue (on which the Be doublet is located), green
and red wavelengths.  The quantum efficiency near 3100 \AA\ is about 93\%,
making exposures of stars down to V = 12 possible.

Stars were selected based on their metallicity and observability.  Stars with
[Fe/H] values of less than $-$2.4 dex were selected from a literature search.
To obtain the best spectra, stars were observed near the meridian, at the
lowest possible airmass and atmospheric dispersion.  Our aim was to observe
stars with a declination range of $-20 < \delta < +60$; the latitude of Mauna
Kea is +19$^{\deg}$ 45'.  In order to obtain spectra of multiple objects with
the required S/N, a cutoff at V of 12.5 was chosen.  Table 1 shows
observations of the stars analyzed with V magnitudes, literature values of
[Fe/H], the UT dates of the observations, the total exposure times and S/N per
pixel near the Be doublet; in the cases where there are exposures on multiple
nights the exposure times and S/N values are the totals.  The median S/N = 104
and the mean is 106 per pixel.

HIRES data reduction was carried out using the IDL HIREDUX pipeline and
standard IRAF routines.  The pipeline was used to bias-subtract,
flatten, extract and wavelength calibrate the spectrum from each individual
exposure.  IRAF was used to co-add multiple exposures of the same star and to
fit continua to each combined, calibrated spectrum.  G 64-37 was observed with
HDS on Subaru, which has a similar but smaller wavelength range, with a
maximum of $\sim$ 4500 \AA; see Boesgaard \& Novicki (2006).

In addition to the new data obtained with the upgraded HIRES we have included
data from Boesgaard et al.~(1999a), as reanalyzed here, data from the Subaru 8
m telescope and HDS by Boesgaard \& Novicki (2006) and data from Keck I with
HIRES for three standard stars by Boesgaard (2007).

\section{Data Analysis}

The data were analyzed using IRAF and MOOG, a stellar synthesis and analysis
program\footnote{http://verdi.as.utexas.edu/moog.html} (Sneden 1973).  We
used the 2002 version, which includes the UV opacity edges of the major atomic
species. Equivalent widths of spectral lines were measured using IRAF
routines.  MOOG was used for two purposes: stellar parameter determination and
abundance determinations.  MOOG's ``abfind'' driver was used with equivalent
widths of Fe I, Fe II, Ti I, and Ti II lines to measure the metallicity,
[Fe/H], and determine $T_{\rm eff}$ and the surface gravity, log $g$, of the
stars in our sample.  We used the ``synth'' driver to create synthetic spectra
in order to measure Be and O abundances.

\subsection{Stellar Parameters}
Stellar parameters are needed to generate the stellar models used for spectral
synthesis and abundance determination.  We have chosen to determine our
stellar parameters spectroscopically.  We are dealing with spectral lines and
need to know the temperature in the region of the atmosphere where the lines
are formed.  These temperatures seem more relevant for abundance work than the
continuum-based color temperatures.  It is generally the case that
spectroscopic temperatures are lower than color temperatures because the
continuum is formed deeper in the stellar atmosphere where the temperatures
are higher.

Several Fe I and Fe II and Ti I and Ti II lines fall within the range of the
HIRES CCD and can be used to determine [Fe/H], $T_{\rm eff}$ and log $g$,
using an iterative method similar to that of Stephens (1999).
Microturbulence, $\xi$, has a negligible effect on any of the measurements in
this project, so a standard value of 1.5 km/s was used in all models (Magain
1984).  We measured equivalent widths of 30-60 Fe I and 10 Fe II lines on the
green and red chips of the HIRES CCD as well as 3 Fe I lines from the UV chip
near the Be doublet.  The Fe I lines covered a range in excitation potential
from 0.86 to 3.98 eV and so could be used to derive the temperature.  In
addition we measured equivalent widths of 10-15 Ti I lines and 8-12 Ti II
lines.  A list of the lines measured, along with the excitation potential and
the $gf$ values, are given in Table 2.  The values for $gf$ come from the
compilation presented in Stephens (1999) and Stephens \& Boesgaard (2002)
which gives the appropriate source references.

We have been careful to use only weak lines which would be on the linear
portion of the curve of growth.  This corresponds to log (W/$\lambda$) $<$
$-$4.82 which is $\sim$75 m\AA\ at 5000 \AA.  For some stars we applied a even
stricter limit of log (W/$\lambda$) $<$ $-$5.15 which is $\sim$35 m\AA\ at
5000 \AA.  The measured equivalent widths are given in the tables in the
Appendix.  For two stars, G 64-12 and G 64-37, we adopted the parameters of
Stephens \& Boesgaard (2002) which were also determined spectroscopically;
they used the same method as we did here, but they had a larger wavelength
coverage toward the red and thus more useable lines.

Using an initial estimate for the stellar parameters and the measured
equivalent widths, MOOG's ``abfind'' driver was then used to calculate an Fe
abundance for each line, as well as an average [Fe/H] from the equivalent
widths.  MOOG also calculates the slope of each line's calculated abundance
versus its excitation potential (EP).  If the temperature is correct, there
should be no trend of abundance with EP.

Atomic abundances derived from two ions of the same species should give
similar results.  If the abundances do not agree, the value of log $g$ can be
adjusted and the abundances recalculated with MOOG.  For the purposes of this
project, MOOG was run iteratively by first calculating the abundance from Ti I
lines using the temperature found via the Fe I lines, then finding the
abundance from Ti II and comparing the two sets of abundances.  The value for
log $g$ is adjusted and the process can then be run iteratively between Fe I
and Ti I plus Ti II abundances until [Fe/H], $T_{\rm eff}$ and log $g$ are
calculated.  (We have used the ionization balance between Ti I and Ti II,
rather than Fe I and Fe II out of concern about NLTE effects on Fe I lines in
metal-poor stars, e.g.~Th\'evenin \& Idiart 1999.)  The stellar parameters
used to calculate abundances are shown in the second, third and fourth columns
of Table 3.  Typical errors are $T_{\rm eff}$ $\pm$80 K, log $g$ $\pm$ 0.2
dex, [Fe/H] $\pm$0.15 dex.  Although our random errors on the temperatures
determined spectroscopically are typically $\pm$80 K, we point out that there
may be large (100-250 K) systematic differences in temperature determinations
by other methods such as UBV and other photometric indices (and the potential
uncertainties in the reddening corrections), the infrared flux method,
Balmer-line profiles.

Although we have relied on the Magain (1984) value of 1.5 km s$^{-1}$ for the
microturbulence, we have checked the effect of using 1.0 km s$^{-1}$ on the
stellar parameters.  On average with $\xi$ = 1.0 km s$^{-1}$, the value of
$T_{\rm eff}$ is lower by 66 K, log $g$ is lower by 0.15 and [Fe/H] is lower
by 0.03 dex.  All of these are within the range of the uncertainities in our
parameter determinations.

Ten of our stars have been observed for Li by Hosford et al.~(2009).  They use
the excitation balance of Fe I to constrain their temperature scale and they
constrain log $g$ using theoretical isochrones.  They present two sets of
parameters for most of the stars appropriate for a main-sequence star and a
subgiant-branch star.  Our temperatures for the 10 stars in common are
typically lower than theirs by $-$163 K on average.  Our log $g$ values are
usually more in alignment with their subgiant branch gravities and on average
are lower by $-$0.05, well within the uncertainty of the determination.  While
the Li abundances are sensitive to temperature, but not to log $g$, the Be
abundances are quite insensitive to temperature, but are sensitive to log $g$
(see $\S$3.2 below.)

\subsection{Abundance Determination and Errors}

The difficulty in determining A(Be) in metal poor stars can be seen in Figure
1, which shows a comparison of a very metal poor star, BD +3 740 at [Fe/H] =
$-$2.95 and one of much higher metallicity, HD 194598 at [Fe/H] = $-$1.23,
from Boesgaard (2007).  Both Be II lines are considerably weaker in the low
metallicity star, BD + 3 740, and high signal-to-noise ratios are needed in
spectra from such stars.  The blending lines are also weaker, however, and the
continuum placement is easier in the metal poor stars.

Once stellar parameters were calculated for each star, they were used to
interpolate models from the Kurucz (1993) grid.  The models, coupled with a
line list (adapted from Boesgaard et al.~1999a) generate a synthetic spectrum,
which can then be compared to the data.  The data are shifted in wavelength
space to match the model and gaussian smoothing is applied to the model
comparable to the shape of the line.  Residuals of the synthetic spectrum
versus the data can be viewed in MOOG in order to determine appropriate values
for the wavelength shift and gaussian smoothing.  The synthesis of the Be
region of two stars is shown in Figure 2.  Both Be and O abundances are
optimized for the best fit in Figure 2.

The Be abundances determined this way are listed in Table 3 for each star.

Inasmuch as Li is more fragile with respect to nuclear destruction than Be, we
looked at Li abundance determinations to ascertain whether there might be Be
deficiencies.  All of our stars except one (LP 752-17) have been observed for
the abundance of Li, and all have normal Li-plateau abundances.  Based on the
criterion that normal Li-plateau values imply normal Be abundances, we
conclude that none of these 23 stars has undergone Be depletion or dilution.
Eleven of our stars appear in the Li compilation of Ryan et al.~(1996) and
nine are in the study of Ryan et al.~(1999).  The mean A(Li) = 2.17 $\pm$0.12.

Although O is one of the most abundant elements in the universe, the
determination of the O content in stars is not straight-forward.  The three
spectral features most commonly used are the OH lines in the UV, the O I
triplet near 7774 \AA\, and the [O I] line at 6300 \AA.  All three features
present drawbacks, see e.g. Boesgaard (2001), Asplund \& Garc\'\i a P\'erez
(2001).  In this work we are constrained to use the OH features in the UV
which are near the Be II resonance lines.  Our spectra do not extend past 6000
\AA.  Furthermore, the [O I] line is very weak in metal-poor stars, and in
dwarfs and subdwarfs in particular; the O abundances found from [O I] for red
giants may reflect altered O abundances due to nuclear processing of O along
with mixing to the stellar surface.

There are electronic transition lines of OH that fall on the UV chip which we
used to calculate the O abundance of each star using the ``synth'' driver as
well.  Some of these lines fall within the Be synthesis region, just redward
of the Be line at 3130 \AA; two others region are further toward the red at
3139 \AA\ and 3140 \AA.  The line list for the 3139 region covers 1.5 \AA\ and
contains seven OH lines, the strongest of which is at 3139.17 \AA\ with a log
$gf$ of $-$1.71, as well as atomic lines.  The line list in the 3140 region
covers 1.1 \AA\ and five OH transitions plus atomic lines.  Generally, the O
abundance agreed with the initial abundance calculated in the Be region.  The
average values are listed in Table 2, where the 3130 \AA\ line has 2 times the
weight of the other two lines.  The synthesis fits for the two other OH line
regions that we used are shown in Figure 3 for HD 140283 which has [Fe/H] =
$-$2.56.

Errors in the abundance determinations are a result of error in the stellar
parameters as well as the accuracy of the synthesis fits.  Errors in the
abundances were determined by adjusting the stellar parameters [Fe/H], $T_{\rm
eff}$ and log $g$ to see how it affected the abundance calculation.  The
largest effect on the Be abundances of any of the stellar parameters is due to
uncertainties in log $g$.  For instance, a change in $T_{\rm eff}$ in $\pm$80
K results in a change in A(Be) of $\pm$0.02 and a change in [Fe/H] of $\pm$0.2
dex results in a change in A(Be) of $\pm$0.01.  However, a change in log $g$
of $\pm$0.2 results in a change of about $\pm$0.10 dex in A(Be).  These errors
are added in quadrature, which results in a typical error in A(Be) of about
$\pm$0.12 dex.  The individual errors are given in Table 3.

The errors on O abundances also result from the uncertainties in the stellar
parameters and were determined in a similar manner.  For the OH features,
however, the error in $T_{\rm eff}$ has the largest effect on the determined
abundance, followed by log $g$.  A change in $T_{\rm eff}$ of $\pm$80 K
changes [O/H] by $\pm$0.15 to 0.17 dex, while a change in log $g$ of +0.2 only
results in a change of [O/H] by +0.05 to +0.07 dex.  These uncertainties were
added in quadrature and appear in Table 3.

Although the ``missing'' UV opacity has been blamed for inaccuracies in the Be
abundances (Balachandran \& Bell 1998), in these stars of such low
metallicity, there is no problem of continuum placement and no issue of
missing opacity.  Similarly, the O abundances from the UV lines of OH are
basically unaffected by opacity or continuum placement issues.  We point out
again that the 2002 version of MOOG includes the UV opacity edges of the
dominant elements.  In considering this issue Smiljanic et al.~(2009)
calculated the effect of an increase of 0.20 in [Fe/H] (a factor of 1.6 as
suggested by Balachandran \& Bell 1998) on Be for a star with [Fe/H] = $-$0.5;
they found that the difference in A(Be) was a negligible effect of only 0.022
dex.

Another source of uncertainty in the O abundances found from the OH features
is in our use of standard LTE 1-D hydrostatic model atmospheres.  Asplund \&
Garc\'\i a P\'erez (2001) use 3-D hydrodynamical model atmospheres to assess
the effect of their more sophisticated models on the O abundance.  They find
that the corrections to [O/Fe] increase with decreasing [Fe/H] and also
increase with decreasing gravity ranging from 0.3 to 0.4 dex at [Fe/H] =
$-$2.0 to 0.5 to 0.6 dex at [Fe/H] = $-$3.0.  They suggest that some of the
calculated decrease in O may be mitigated by a full calculation of non-LTE
effects.

We note that the absolute solar O abundance does not affect our [O/H] values
as the MOOG input value is just subtracted out.  The same is true for the Be
abundance; it does not matter whether we use the meteoritic Be abundance or
some solar abundance.

\subsection{New Be and O abundances from Previous Observations}

We have reanalyzed the Keck/HIRES observations in Boesgaard et al.~(1999a,
1999b) to use the newer version of MOOG from 2002.  These observations were
made with the original HIRES and due to the limited wavelength coverage of
that CCD, we could not determine the stellar parameters spectroscopically.
Instead we used the the parameters from the 1999 papers with temperatures on
the Carney (1983) scale.  That temperature scale is consistent with the
spectroscopic temperatures derived for the stars in this paper.  For three
stars we have lowered the temperature and achieve better fits.  The new Be
abundances were found by the same method used here - spectrum synthesis - and
with the same line list.  We have rederived the O abundances but use only
three OH features (3130, 3139, and 3140 \AA), as was done here for the newly
observed stars.  The new abundance results and the 1$\sigma$ error estimates
are presented here in Table 4 along with the stellar parameters and the
1$\sigma$ errors on those.

The Be abundances do not differ much from those presented in Boesgaard et
al.~(1999a): on average they differ by $-$0.005 $\pm$0.077 with a range in
$\Delta$A(Be) of $-$0.14 to +0.15 (in the sense of new minus old).  The O
abundances differ systematically when we use just the three clean OH features.
The difference in $\Delta$[O/H] ranges from $-$0.55 to +0.12 and the average
is $-$0.22 $\pm$0.16 (again in the sense of new minus old).  The differences
result from our use of the three best OH features and ignoring the other weak
and blended OH lines used in the earlier study.

\subsection{Be and New O Abundances for Published Metal-Poor Stars}

We have incorporated in this study the results on Be in metal-poor stars
published by Boesgaard \& Novicki (2006) (four stars), and the three Li-Be
normal stars from Boesgaard (2007).  These stars, their parameters and
abundances are given in Table 4.  We have determined O abundances in these
stars in the same manner as above, from 3 OH features.  Those results are
presented in Table 5.

\subsection{Be Abundance Comparisons}

There are two stars in common in this study with Boesgaard et al.~(1999a): BD
+3$^{\arcdeg}$ 740 and BD $-$13$^{\arcdeg}$ 3442.  For BD +3$^{\arcdeg}$ 740
the parameters determined spectroscopically and photometrically agree well.
The temperatures differ by $-$80 K, log $g$ by 0.21, [Fe/H] by $-$0.06, [O/H]
by $-$0.10, and A(Be) by $-$0.03 (in the sense of this study minus the
reanalyzed (Table 3) values from the data from Boesgaard et al.~1999a).  In
the case of BD $-$13$^{\arcdeg}$ 3442 the spectroscopic gravity determined
here is considerably higher (4.11 versus 3.50) and thus A(Be) is lower by 0.6
dex.  The low value for log $g$ from Boesgaard et al.~(1999a) is very
uncertain ($\pm$0.44 in Table 3) so we adopt the results from this study.  In
addition, the S/N of the spectrum used here is 183 compared to the earlier one
at 129.

There are two stars analyzed in this project that were previously considered
by Primas et al.~(2000a, 2000b), LP 815-43 and G 64-12.  The values of A(Be)
calculated from our data are slightly different from theirs: for LP 815-43 we
derived A(Be)= $-$0.95, while Primas et al.~(2000a) found A(Be)= $-$1.09 and
for G 64-12 we found A(Be)= $-$1.43, whereas Primas et al.~(2000b) derived
A(Be)= $-$1.15.  Primas et al.~(2000a, 2000b) adopted different stellar
parameters for their stellar models, which would lead to different abundances.
The parameters used in this project were determined spectroscopically and fit
the data well.  When the stellar parameters from Primas et al.~(2000a) were
applied to our data for LP 815-43, A(Be) was found to be about $-$1.10, but
other spectroscopic features in the Be region did not fit the data as well as
with their parameters as with ours.

Although there are no stars from Smiljanic et al.~(2009) in common with our
new Keck data set in Table 3, there are 11 in common with our revised
Boesgaard et al.~(1999a) values in Table 4.  Even though they used log $g$
values derived from parallax and ours are from ionization balance, with the
exception of one star (HD 219617) the agreement in log $g$ is good with a mean
difference of +0.01 $\pm$0.20 (in the sense of this study minus Smiljanic et
al.).  The differences in A(Be) range from $-$0.17 to +0.19 dex (in the sense
of this study minus Smiljanic et al.).  The mean difference is +0.02 $\pm$0.13
dex, indicating agreement within the individual errors.  Two of the seven
stars in Table 4 were also studied by Smiljanic et al.  For HD 194598 our
value for A(Be) is +0.05 dex higher and for HD 195633 our value is +0.18 dex
higher.

There are four stars in common from Tables 4 and 5 with Tan et al.~(2009): HD
76932, HD 132475, HD 140283, and HD 195633.  The mean difference in our values
minus theirs for A(Be) is $-$0.03 $\pm$0.16 dex with the largest difference
($-$0.24 dex) being that for HD 140283 caused by stellar parameter
differences.  

Interestingly, there are nine stars in common to the studies of Tan et
al.~(2009) and Smiljanic et al.~(2009) from the same ESO/UVES data set.  Their
agreement for A(Be) is good with a mean difference (Smiljanic minus Tan) of
+0.07 $\pm$0.09 dex.

\section{Results and Discussion}
In this paper we discuss the abundance relationships between Be and Fe,
between Be and O, and between Fe and O.  We have acquired additional
observations of Be in 34 more stars with [Fe/H] between $-$0.7 and $-$2.5.  A
paper by Boesgaard et al.~(in preparation) will address the issue of the
possibility of a spread in Be at a given [Fe/H] and [O/H] in the larger sample
of 82 stars.  That larger sample will also examine the potential connection
between A(Be) and [O/Fe] and the use of Be as a chronometer.  Preliminary
results from a smaller sample on these matters have appeared in a conference
proceeding by Boesgaard et al.~(2008).

\subsection{Be and Fe Relationships}
The Be abundances from Tables 3, 4, and 5 are shown in Figure 4 plotted
against the values of [Fe/H].  The abundances show a trend of increasing A(Be)
with increasing [Fe/H].  

The straight line between A(Be) and [Fe/H] is

A(Be) = 0.92 ($\pm$0.04)[Fe/H] + 1.41 ($\pm$0.09).

One of our stars in Table 3, LP 831-70, has only an upper limit on A(Be); it
is plotted as an inverted triangle in the figures and it has not been included
in the linear fits.  The slope of the least squares fit is 0.92 $\pm$0.04.
This slope is similar to that found by Boesgaard et al.~(1999a) of 0.96
$\pm$0.04.  It is in agreement with the value of Molaro et al.~(1997) of 1.07
$\pm$0.07, but less steep than the recent values of 1.16 to 1.27 $\pm$0.07
found for the samples in Smiljanic et al.~(2009).  Tan et al.~(2009) augmented
their sample with the results for A(Be) and [Fe/H] from Boesgaard et
al.~(1999a) in their Figure 7; considering their sample of 21 stars (omitting
four ``abnormal'' Be values) they derive a slope of 1.10 $\pm$0.07.

In Figure 5 we show two fits: one with a slope of 0.56 $\pm$0.07 for the stars
with [Fe/H] $<$$-$2.2 and one of 1.07 $\pm$0.06 for stars with [Fe/H]
$>$$-$2.2.  The slope for the higher [Fe/H] stars is in good agreement with
those reported in the earlier studies, most of which contain only stars with
[Fe/H] $>$ $-$2.5.  The shallower slope for the metal-poorest stars may
indicate a slower, but gradual increase in Be as Fe increases in halo stars.
On the other hand, the fit with one line is statistically no better or worse
than the one with two lines.  We have also tried making the break at [Fe/H] =
$-$2.7; the fit is less good and the slopes are nearly the same at 0.57
$\pm$0.12 and 1.05 $\pm$0.05.  The two slopes intersect at $-$2.3, not $-$2.7.

The possible nucleosynthesis of Be in the Big Bang has been discussed by
several authors (e.g.~Boyd et al.~1989, Malaney and Fowler 1989, Alibes et
al.~2002, Jedamzik and Rehm 2006).  Only in some inhomogeneous BB cosmologies
is there enough Be produced to observe using current instruments.  Our results
do not seem to indicate a Be plateau at very low metallicities, as has been
suggested by Primas et al.~(2000a).  Analyzing stars of even lower
metallicities might provide further information, but observations become very
time consuming, due to the faintness of the extremely low metal stars and the
need for high S/N in order to observe the increasingly weak Be II lines.  Our
results in Figures 4 and 5 show no evidence for a plateau of constant A(Be),
but a slower increase in Be with Fe at very low metallicities.

A more sensitive way to search for evidence of a plateau is to examine the Be
abundance normalized to Fe, shown in Figure 6.  The four stars with [Fe/H] $<$
$-$3.0 do show enhanced Be relative to Fe.  The best fit relationship is

[Be/Fe] = $-$0.08 ($\pm$0.04) [Fe/H] $-$0.01 ($\pm$0.09).

The dashed line in the figure corresponds to the mean value of [Be/Fe] for the
48 stars which is +0.17 $\pm$0.21.  This probable error bar is shown at the
left in Figure 6 along the mean value line for [Be/Fe].  Although there is
considerable scatter in [Be/Fe] (the range is $-$0.21 to +0.65), the stars
with the highest values of [Be/Fe] are generally the lowest metallicity stars.
But, as in Figures 4 and 5, there is not a plateau with a constant A(Be) or
[Be/Fe], but rather a smooth trend.  The removal of the two Be-rich stars, HD
94028 and HD 132475 near [Fe/H] of $-$1.5, reduces the mean only to +0.16 but
shows a clearer trend of high [Be/Fe] at low [Fe/H].  (We note that Smiljanic
et al.~2009 found a lower log $g$ for HD 94028 and thus a lower Be abundance,
putting it in alignment with the other stars.)

\subsection{Be and O Relationships}

The nucleosynthesis of Be is directly related to O as the spallation of
abundant atoms like C, N, and O produce smaller atoms like Li, Be, and B.  The
spallation can occur 1) as high energy cosmic rays bombard CNO atoms in the
ambient interstellar gas or 2) during supernovae explosions where atoms of CNO
and protons are accelerated to high energies in the vicinity.  These two
mechanisms predict different relationships between Be and O.  

In the more traditional GCR spallation the slope between A(Be) and [O/H] would
be 2 because the number of O atoms is proportional to the cumulative number of
SN IIa ($N$) while the energetic protons are proportional to the instantaneous
number of SN II ($dN$).  The integral of $N dN$ is $kN^2$.

In the immediate vicinity of SN II the number of Be atoms would be
proportional to the number of O atoms; this would result in a slope of 1
between A(Be) and [O/H].

In Figure 7 we show the least-squares fit between A(Be) and [O/H].  The
equation for the straight line is

A(Be) = 1.21 ($\pm$0.08) + 1.32 ($\pm$0.14).

There are reasons to suspect that Galactic evolution will play a role in the
rate of the production of SN II.  The early generations of stars will be
massive stars which will produce supernova and begin to enrich the next
generations of stars.  These early SN will expel C and O which has been
produced by nuclear fusion in their interiors prior to the supernova phase.
The dominant mechanism creating Be atoms would thus be \#2 above: acceleration
of CNO from the SN explosion.  Eventually, after multiple generations of stars
have formed, the dominant mechanism would be \#1 above: spallation by high
energy cosmic rays in the interstellar gas.

In Figure 7 one can discern a possible change in slope so we have tried
fitting two lines to the data.  This result is shown in Figure 8.  The oldest
stars with the least O, [O/H] $<$ $-$1.4, have a shallower slope of 0.74
$\pm$0.11, while the stars with [O/H] $>$ $-$1.8 have a steeper slope of 1.59
$\pm$0.15.  The transition occurs near [O/H] of $-$1.6.

For the stars with the lowest [O/H] the relationship is

A(Be) = 0.74 ($\pm$0.11) [O/H] + 0.40 ($\pm$0.22).

For the stars with the higher values of [O/H] the relationship is

A(Be) = 1.59 ($\pm$0.15) [O/H] + 1.75 ($\pm$0.20).

A changing slope is consistent with the change expected as the dominant
production method for the spallation of Be changes.  Other effects enter into
the determination of the relationship between Be and O, such as mass outflow
which may be lost during the process of star formation.  Such effects would
modify the predicted slopes of 1 and 2 to lower values.  While we do not plan
to discuss the models of Galactic chemical evolution, we point to these
observations as a potential constraint on such models.

With respect to the [O/H] abundances in Figures 7 and 8, we are concerned
about the reduction in the O abundances through the use of 3-D hydrodynamic
models as discussed by Asplund \& Garc\'\i a P\'erez (2001).  The decrease in
O becomes larger as the metallicity decreases.  The stars with the lowest
values of A(Be) would have decreased [O/H] values.  The low Be points in those
two figures would move to the left along the x-axis, thus favoring the 2-slope
solution.

\subsection{Fe and O Relationships}

Our data enable us to reenter the controversy regarding the form of the
relation between O and Fe.  Our sample contains 49 stars which are all dwarfs
and subgiants - no giants; the range in log $g$ is 3.20 to 4.52.  The
temperatures range from turn-off stars at 6400 K to dwarf stars at 5500 K.
Our star sample covers 3 orders of magnitude in [Fe/H] from $-$0.5 to $-$3.5
with the bulk of the stars, 34, being below $-$2.0 and six of them $\leq$
$-$3.0.  This is a large uniform collection of similarly-analyzed stars which
pushes the boundary to lower metallicities in an unevolved sample than was
done, for example by Nissen et al.~(2002), Fulbright \& Johnson (2003),
Garc\'\i a P\'erez et al.~(2006).  Although the Cayrel et al.~(2004) study
does go to low values of [Fe/H], their stars are low-gravity evolved stars
with log $g$ ranging from 0.7 to 2.7; they find a mean [O/Fe] of 0.7 with a
dispersion of 0.17 dex and 0.47 after an uncertain correction for stellar
surface inhomogeneities.

In Figure 9 we show the tight, linear relationship between [O/H] and [Fe/H]
with a well-fitting slope of 0.70 $\pm$0.03.  The lowest metallicity stars
merge well into the stars with higher values of [Fe/H].

The relationship is

[O/H] = 0.692 ($\pm$0.032) [Fe/H] $-$ 0.076 ($\pm$0.074)

Figure 10 shows the usual relationship where the element is normalized to Fe,
in this case [O/Fe].  This can be fit well with a linear relationship where
[O/Fe] is $\sim$1 at [Fe/H] = $-$3.5 and progresses towards 0 at [Fe/H] = 0.
The equation is

[O/Fe] = $-$0.30 ($\pm$0.03) [Fe/H] $-$0.05 ($\pm$0.07)

The continued upward trend toward lower metallicities in this relationship is
similar to that found by Israelian et al.~(1998, 2001), Boesgaard et
al.~(1999b), Nissen et al.~(2002) although none of those studies went to as
low values of [Fe/H] as in this work.  Fulbright \& Johnson (2002) show a very
similar trend with their ad hoc temperature which forces the parameters to
yield agreement between the permitted and forbidden O features (see their
Figure 13).  

\subsection{Implications}

The linearity of the relation between [O/H] and [Fe/H] in Figure 9 is
striking.  And it implies that the 2 slope fit between A(Be) and [O/H] (Figure
8) probably requires the 2 slope fit between A(Be) and [Fe/H] (Figure 5).  In
turn this indicates that the upturn in [Be/Fe] in Figure 6 at the lowest
metallicities is significant.  It is not evidence for a ``plateau'' in the Be
abundance, but rather a change in the chemical history of the formation of Be.
As discussed in $\S$4.2 the spallation reactions occurring in the vicinity of
SN II would dominate in the early episodes of Galactic evolution and there
would be a flatter slope between Be and O.  While still continuing to produce
Be by this means, the standard GCR spallation would be more dominant in later
eras, thus a steeper slope between Be and O.  Again, we note that other
stellar formation and evolution effects enter into the predicted relation and
reduce both slopes.

\section{Summary and Conclusion}
We have used the upgraded HIRES on the Keck I 10m telescope to obtain high
resolution, high signal-to-noise spectra of 24 very metal deficient stars from
[Fe/H] = $-$3.45 to $-$2.32.  Of the 24 stars 19 have [Fe/H] $<$$-$2.5 and the
other 5 are near $-$2.3.  (Even the large Be sample of Smiljanic et al.~(2009)
contains only three stars in our metallicity range, and those are at the high
end of our range, with [Fe/H] = $-$2.30, $-$2.35, and $-$2.48 and the latter
has S/N of only 45.)  Our data set was supplemented by spectra of 25 stars
from Boesgaard et al.~(1999a), Boesgaard \& Novicki (2006) and Boesgaard
(2007).  For all these stars Be abundances determined from the Be II doublet at
3130 and 3131 \AA\ and O abundances were found from 3 features of OH in the UV
spectral region.  For the newly observed stars stellar parameters were found
spectroscopically using equivalent widths of Fe I for $T_{\rm eff}$, and Ti I
and Ti II for log g.  An iterative procedure was followed for the parameter
determination and the Fe abundance.  The stellar synthesis program MOOG was
used in both {\it abfind} and {\it synth} modes to find the elemental
abundances of Be, O, and Fe.

The full sample of 49 stars reveals a tight correlation between [O/H] and
[Fe/H] with a slope of 0.69 $\pm$0.03 over 3 orders of magnitude in [Fe/H]
from $-$0.5 to $-$3.5 and over 2 orders of magnitude in [O/H] from $-$0.4 to
$-$2.5.  This is shown in Figure 9.  One of the implications of this linear
fit is that the stellar parameters and the abundance determinations give
consistent results for the two sets of stars: the 24 very metal-poor stars
(with spectroscopically determined parameters) and the reanalysis of the data
from the three earlier papers (with photometrically-derived temperatures).

For one star we could only determine an upper limit on the Be abundance so
our sample size is 48 for the Be data.  Of the 48 stars 34 are below $-$2.0 in
[Fe/H] and five of those are $\leq$$-$3.0.  This sample by far the largest
sample of Be abundances found for low metallicity stars.  In the sample of
Smiljanic et al.~(2009) of 73 Be detections, only four have [Fe/H] $<$ $-$2.0
and four more at $-$2.0.

We derive a linear relationship between A(Be) and [Fe/H] which has a slope of
0.92 $\pm$0.04 and a linear relationship between A(Be) and [O/H] with a slope
of 1.21 $\pm$0.08.  Both plots can also be well fit by two linear
relationships.  The tightness of the relationship between the two abscissae,
[Fe/H] and [O/H], lends support for the two-slope fits with A(Be).  The slope
change between A(Be) and [O/H] from 0.74 $\pm$0.11 at low [O/H] to 1.59 $\pm$
0.15 occurs at [O/H] = $-$1.6.  See Figure 8.  The corresponding change point
occurs at [Fe/H] = $-$2.25.  See Figure 5.  This ``kink'' occurs near $-$0.9
for A(Be).

The change in slope between A(Be) and [O/H] would be a natural consequence of
the change in the dominant mechanism of the formation of Be.  In the early
eras of the Galaxy, when massive stars become supernovae, the Be would be
formed predominantly in the vicinity of the SN II where CNO atoms are
accelerated to high energies and break up as they bombard protons, for
example, and form lighter atoms such as Be.  The slope would be a maximum of
1.0 as we find 0.74 for the stars with the lowest values of [O/H].  Later the
spallation reactions could occur as high energy cosmic rays bombard CNO atoms
in the interstellar gas.  In that case the slope would be a maximum of 2.0,
where we find 1.59.  The change between the two mechanism's dominance occurs
as [O/H] increases from $-$1.8 to $-$1.4.

As we have found, the relationship between [O/H] and [Fe/H] is linear; the
change in the slope between A(Be) and [Fe/H] also results from the change in
the dominant mechanism that forms Be.  Whereas the formation of Be and O are
related, the same cannot be said of Be and Fe.  The Fe atoms are a product of
stellar nucleosynthesis via the e-process and not directly connected to the
formation of Be.  The slope change is due to Be production, not Fe.

One of our goals was to search for a Be plateau at low metallicities,
analogous to the Li plateau.  We searched for this in our plot of [Be/Fe] vs
[Fe/H] in Figure 6.  We did not find evidence for a constant Be abundance at
the lowest values of [Fe/H], but rather an indication of a gradual increase in
[Be/Fe] toward the lowest [Fe/H] values.  As Figure 6 shows, the four stars
with the lowest [Fe/H] have enhanced [Be/Fe] equivalent to the one sigma
error.  These data do not indicate that Be has been formed in an
inhomogeneous-style Big Bang.

We have a large sample of low metal stars in which O and Fe abundances have
been determined consistently so we can use the connection between [O/Fe] and
[Fe/H] to learn about the early overproduction of O (formed in massive stars)
relative to Fe (formed in intermediate-mass stars).  We looked at the
enhancement in [O/Fe] as a function of [Fe/H] in Figure 10.  We find that the
O enhancement decrease steadily from [O/Fe] near $-$1.0 at [Fe/H] near $-$3.5
to 0.0 near [Fe/H] = 0.0.  The slope of the fit between the two parameters is
$-$0.30 $\pm$0.03.  The [O/Fe] enhancement diminishes steadily rather than
showing a constant enhancement below [Fe/H] of $-$2.0 of 0.7 $\pm$0.17 as
found by Cayrel et al.~(2004).

\acknowledgements We are grateful to the various Support Astronomers and
Observing Assistants who helped us over the 11 observing runs for this
project.  We acknowledge support from NSF through grant AST 05-05899 to A.M.B.

\clearpage
\tightenlines
\singlespace
\centering
\begin{deluxetable}{lrllclcl} 
\footnotesize
\tablewidth{0pc}
\tablecolumns{8}
\tablenum{1} 
\tablecaption{Log of the Beryllium Observations} 
\tablehead{ \colhead{Star}  &  \colhead{Code\tablenotemark{a}} &
\colhead{R.A.}  
& \colhead{Dec.}  & \colhead{V} & \colhead{Date(s) U.T.} & \colhead{Exp. min.} & 
\colhead{S/N} 
}

\startdata
LP 651-4     & 1  & 02 44 &  $-$05 26  &  12.04  & 2006 Jan 2 & \phn 65  & \phn 83 \\  
G 75-56      & 2  & 03 00 &  $-$05 57  &  11.94  & 2004 Sep 7 & & \\
	     &    &	&            &         & 2004 Nov 7 &  232   & 107 \\	  
LP 831-70    & 3  & 03 06 &  $-$22 19  &  11.40  & 2008 Jan 16 & \phn 90  & \phn 52 \\
BD +3 740    & 4  & 05 01 &+04 06  & \phn 9.80   & 2004 Nov 7 &  \phn 60  & 166 \\  
G 108-58     & 5  & 07 10 &  $-$01 17  &  11.82  & 2008 Jan 16 &  \phn 90 & \phn 91 \\
G 88-10      & 6  & 07 10 &    +24 20  &  11.86  & 2007 Nov 19 & & \\
             &  &	&            &         & 2008 Jan 16 &  150   &  121 \\
BD +20 2030  & 7  & 08 16 &    +19 41  &  11.20  & 2006 Jan 2 &  \phn 60  & 122 \\ 
BD +9 2190   & 8  & 09 29 &    +08 38  &  11.16  & 2005 Jan 31 &  120   & 110 \\	   
BD +1 2341p  & 9  & 09 40 &    +01 00  &  10.48  & 2006 Jan 2 &  \phn 40  & 139 \\  
BD +44 1910  & 10 & 09 49 &    +44 17  &  10.95  & 2008 Jan 16 &  \phn 60 & 112 \\
BD $-$13 3442  & 11 & 11 46 &$-$14 06  &  10.37  & 2006 Jan 2 &  \phn 80  & 194 \\  
LP 553-62    & 12 & 11 54 &    +02 57  &  11.68  & 2008 Jan 16 &  \phn 90 & \phn 97 \\
G 11-44      & 13 & 12 10 &    +00 23  &  11.08  & 2008 Jan 16 &  \phn 30 & \phn 76 \\
G 59-24      & 14 & 12 34 &    +15 16  &  12.02  & 2008 Jan 16 &  \phn 56 & \phn 73 \\
G 64-12      & 15 & 13 40 &  $-$00 02  &  11.49  & 2005 May 15 &  210   & 109 \\	  
G 64-37\tablenotemark{b}      &    & 14 02 &  $-$05 39  &  11.14  & 2003 May 27 & 270 & \phn 87 \\  
G 201-5      & 16 & 14 36 &    +55 33  &  11.52  & 2007 Jun 10 &  120 &   115 \\
BD +26 2621  & 17 & 14 54 &    +25 33  &  11.00  & 2007 Jun 10 &  \phn 60 & 123 \\
G 206-34     & 18 & 18 35 &    +28 41  &  11.40  & 2005 May 15 &  105   & \phn 97 \\	  
LP 752-17    & 19 & 19 25 &  $-$11 56  &  11.88  & 2005 Jul 6 & & \\
             &   &	 &            &         & 2005 Sep 27 & & \\
             &   &	 &            &         & 2006 Jun 19 &  240  & \phn 94 \\
G 92-6       & 20 & 19 29 &    +01 01  &  11.75  & 2007 Jun 10 &  120   & 100 \\
LP 635-14    & 21 & 20 26 &  $-$00 37  &  11.33  & 2005 Jul 6 & & \\
             &   &	 &            &         & 2007 Jun 10 &  162   & 114 \\
LP 815-43    & 22 & 20 38 &  $-$20 25  &  10.91  & 2004 Sep 7 & & \\ 
             &   &	 &            &         & 2006 Jun 19 &  120  & \phn 98 \\
G 275-4      & 23 & 23 07 &  $-$23 52  &  12.18  & 2005 Jul 5  & & \\
             &   &	 &            &         & 2005 Jul 6  & & \\
             &   &	 &            &         & 2005 Sep 27 &  220   & \phn 61 \\
\enddata
\tablenotetext{a}{Code is an ID number for the star in the Appendix Table of
equivalent widths.}
\tablenotetext{b}{G64-37 observed on Subaru with HDS.  See Boesgaard \& Novicki (2006).}

\end{deluxetable}

\clearpage
\tightenlines
\singlespace
\centering


\begin{deluxetable}{cccc}
\small
\tablecolumns{4}
\tablewidth{0pc}
\tablenum{2}
\tablecaption{Lines Measured}
\tablehead{
\colhead{Element} &
\colhead{$\lambda$($\rm{\AA}$)} &
\colhead{Ex. Pot.(eV)} &
\colhead{gf} 
}

\startdata
 \ion{Fe}{1} & 3100.304 & 0.99   & 0.1348963  \\
             & 3100.665 & 0.95   & 0.1361445  \\
             & 3116.631 & 1.01   & 0.0221309  \\
             & 4494.563 & 2.1980 & 0.0725270  \\
             & 4528.614 & 2.1760 & 0.1297180  \\
             & 4531.148 & 1.4850 & 0.0074470  \\
             & 4556.126 & 3.6030 & 0.1633050  \\
             & 4592.651 & 1.5580 & 0.0035030  \\
             & 4602.941 & 1.4850 & 0.0061940  \\
             & 4647.434 & 2.9490 & 0.0467200  \\
             & 4678.846 & 3.6030 & 0.1792670  \\
             & 4691.411 & 2.9910 & 0.0326210  \\
             & 4733.591 & 1.4850 & 0.0010300  \\
             & 4736.773 & 3.2110 & 0.1794730  \\
             & 4871.318 & 2.8660 & 0.4111500  \\
             & 4872.137 & 2.8820 & 0.2609160  \\
             & 4890.750 & 2.8600 & 0.4036454  \\
             & 4891.490 & 2.8400 & 0.7744618  \\
             & 4918.994 & 2.8660 & 0.4405550  \\
             & 4920.503 & 2.8330 & 1.1587773  \\
             & 4985.253 & 3.9290 & 0.2760580  \\
             & 4985.547 & 2.8650 & 0.0466660  \\
             & 4994.130 & 0.9150 & 0.0010740  \\
             & 5001.862 & 3.8820 & 1.0232930  \\
             & 5006.119 & 2.8330 & 0.2298790  \\
             & 5012.068 & 0.8590 & 0.0023820  \\
             & 5022.236 & 3.9840 & 0.2951210  \\
             & 5049.819 & 2.2790 & 0.0447200  \\
             & 5051.635 & 0.9150 & 0.0017220  \\
             & 5068.766 & 2.9400 & 0.0731980  \\
             & 5079.224 & 2.1980 & 0.0082040  \\
             & 5079.740 & 0.9900 & 0.0005850  \\
             & 5083.339 & 0.9580 & 0.0014390  \\
             & 5098.697 & 2.1760 & 0.0094190  \\
             & 5123.720 & 1.0110 & 0.0008650  \\
             & 5150.840 & 0.9900 & 0.0009550  \\
             & 5171.596 & 1.4850 & 0.0174985  \\
             & 5192.344 & 2.9980 & 0.3793150  \\
             & 5194.942 & 1.5580 & 0.0088000  \\
             & 5198.711 & 2.2230 & 0.0077090  \\
             & 5202.336 & 2.1760 & 0.0139800  \\
             & 5215.182 & 3.2660 & 0.1345860  \\
             & 5216.274 & 1.6080 & 0.0076560  \\
             & 5217.390 & 3.2110 & 0.0765600  \\
             & 5227.190 & 1.5570 & 0.0592925  \\
             & 5232.940 & 2.9400 & 0.8016780  \\
             & 5242.491 & 3.6350 & 0.1248820  \\
             & 5250.646 & 2.1980 & 0.0076740  \\
             & 5263.305 & 3.2660 & 0.1189870  \\
             & 5269.537 & 0.8590 & 0.0475883  \\
             & 5328.039 & 0.9150 & 0.0342373  \\
             & 5332.900 & 1.5570 & 0.0013870  \\
             & 5339.930 & 3.2660 & 0.2089300  \\
             & 5341.024 & 1.6080 & 0.0098510  \\
             & 5393.167 & 3.2410 & 0.1539930  \\
             & 5397.128 & 0.9150 & 0.0102920  \\
             & 5405.775 & 0.9900 & 0.0140605  \\
             & 5429.696 & 0.9580 & 0.0131825  \\
             & 5434.524 & 1.0110 & 0.0075510  \\
             & 5455.609 & 1.0110 & 0.0080445  \\
             & 5497.516 & 1.0110 & 0.0014550  \\
             & 5501.465 & 0.9580 & 0.0010050  \\
             & 5569.618 & 3.4170 & 0.3069020  \\
             & 5572.841 & 3.3970 & 0.5099180  \\
             & 5576.090 & 3.4300 & 0.1000000  \\
             & 5586.756 & 3.3680 & 0.7177940  \\
             & 5658.816 & 3.3970 & 0.1458810  \\
\\
 \ion{Fe}{2} & 4508.289 & 2.8580 & 0.0047753  \\
             & 4515.339 & 2.8440 & 0.0033113  \\
             & 4522.634 & 2.8440 & 0.0077625  \\
             & 4576.339 & 2.8410 & 0.0011092  \\
             & 4583.837 & 2.8070 & 0.0120226  \\
             & 4629.339 & 2.8070 & 0.0042658  \\
             & 5197.576 & 3.2310 & 0.0068155  \\
             & 5234.630 & 3.2210 & 0.0061660  \\
             & 5276.002 & 3.2000 & 0.0092257  \\
\\					     
 \ion{Ti}{1} & 4518.023 & 0.8259 & 0.5382698  \\
             & 4527.305 & 0.8130 & 0.3388442  \\
             & 4533.239 & 0.8484 & 3.4040819  \\
             & 4534.778 & 0.8360 & 2.1677041  \\
             & 4535.570 & 0.8259 & 1.3182567  \\
             & 4681.908 & 0.0480 & 0.0966051  \\
             & 4981.732 & 0.8484 & 3.6307805  \\
             & 4991.067 & 0.8360 & 2.7289778  \\
             & 4999.504 & 0.8259 & 2.0230192  \\
             & 5016.162 & 0.8484 & 0.3033891  \\
             & 5020.028 & 0.8360 & 0.4385307  \\
             & 5022.871 & 0.8259 & 0.4187936  \\
             & 5035.907 & 1.4602 & 1.8197009  \\
             & 5036.468 & 1.4432 & 1.5346170  \\
             & 5039.959 & 0.0211 & 0.0741310  \\
             & 5064.654 & 0.0480 & 0.1396368  \\
\\					     
 \ion{Ti}{2} & 4501.272 & 1.1156 & 0.1757924  \\
             & 4501.272 & 1.1156 & 0.1757924  \\
             & 4563.761 & 1.2214 & 0.1096478  \\
             & 4571.968 & 1.5719 & 0.2951209  \\
             & 4589.958 & 1.2369 & 0.0162181  \\
             & 4657.203 & 1.2430 & 0.0058210  \\
             & 4779.985 & 2.0478 & 0.0426580  \\
             & 5129.152 & 1.8918 & 0.0407380  \\
             & 5154.070 & 1.5659 & 0.0120226  \\
             & 5188.680 & 1.5819 & 0.0616595  \\
             & 5226.543 & 1.5659 & 0.0501187  \\
             & 5336.781 & 1.5819 & 0.0216272  \\
             & 5381.018 & 1.5658 & 0.0094406  \\
            
\enddata
\end{deluxetable}

\clearpage
\singlespace
\centering
\begin{deluxetable}{llcccccrc} 
\tablewidth{0pc}
\tablenum{3}
\tablecolumns{9} 
\tablecaption{Stellar Parameters and Abundances} 
\tablehead{ 
\colhead{Star} & \colhead{T$_{\rm eff}$ (K)} & \colhead{log g} 
& \colhead{[Fe/H]} & \colhead{$\sigma$([Fe/H])\tablenotemark{a}}
& \colhead{[O/H]} & \colhead{$\sigma$([O/H])} 
& \colhead{A(Be)} & \colhead{$\sigma$(Be)} 
} 
\startdata
LP 651-4    & 6030 & 4.26 & $-$2.89 & 0.08 & $-$2.04 & 0.14  & $-$1.12 & 0.12\\
G 75-56     & 5890 & 3.83 & $-$2.38 & 0.08 & $-$1.74 & 0.14  & $-$0.84 & 0.12\\
LP 831-70   & 5985 & 4.75 & $-$3.06 & 0.11 & $-$2.45 & 0.13  & $<-$0.90 & 0.12\\
BD +3 740   & 6030 & 3.83 & $-$2.95 & 0.09 & $-$2.26 & 0.14  & $-$1.40 & 0.11\\
G 108-58    & 5865 & 4.03 & $-$2.37 & 0.10 & $-$1.33 & 0.15  & $-$1.12 & 0.12\\
G 88-10     & 5945 & 4.00 & $-$2.61 & 0.07 & $-$1.86 & 0.15  & $-$1.08 & 0.12\\
BD +20 2030 & 5978 & 3.61 & $-$2.77 & 0.10 & $-$2.06 & 0.14  & $-$1.23 & 0.12\\
BD +9 2190  & 6008 & 3.85 & $-$3.00 & 0.09 & $-$2.38 & 0.14  & $-$1.22 & 0.11\\
BD +1 2341p & 6402 & 4.24 & $-$2.67 & 0.11 & $-$1.76 & 0.14  & $-$1.00 & 0.12\\
BD +44 1910 & 5878 & 3.56 & $-$2.64 & 0.08 & $-$1.96 & 0.15  & $-$1.11 & 0.12\\
BD $-$13 3442 & 6090 & 4.11 & $-$2.91 & 0.09 & $-$2.15 & 0.15  & $-$1.12 & 0.12\\
LP 553-62   & 6128 & 3.93 & $-$2.73 & 0.08 & $-$2.02 & 0.14  & $-$1.00 & 0.11\\
G 11-44     & 5820 & 3.58 & $-$2.29 & 0.05 & $-$1.63 & 0.15  & $-$1.04 & 0.11\\
G 59-24     & 6112 & 4.10 & $-$2.32 & 0.10 & $-$1.32 & 0.15  & $-$0.69 & 0.11\\
G 64-12     & 6074 & 3.72 & $-$3.45 & 0.04 & $-$2.24 & 0.15  & $-$1.43 & 0.12\\
G 64-37	    & 6122 & 3.87 & $-$3.28 & 0.05 & $-$2.32 & 0.14  & $-$1.40 & 0.11\\
G 201-5     & 5950 & 4.00 & $-$2.54 & 0.08 & $-$1.97 & 0.15  & $-$1.27 & 0.11\\
BD +26 2621 & 6266 & 4.50 & $-$2.69 & 0.09 & $-$1.96 & 0.13  & $-$0.94 & 0.11\\
G 206-34    & 5825 & 3.99 & $-$3.12 & 0.09 & $-$2.37 & 0.14  & $-$1.20 & 0.11\\
LP 752-17   & 5738 & 3.20 & $-$2.38 & 0.08 & $-$1.80 & 0.17  & $-$0.86 & 0.12\\
G 92-6      & 6115 & 4.79 & $-$2.70 & 0.08 & $-$2.28 & 0.12  & $-$0.91 & 0.11\\
LP 635-14   & 5932 & 3.57 & $-$2.71 & 0.07 & $-$2.00 & 0.12  & $-$1.17 & 0.12\\
LP 815-43   & 6405 & 4.37 & $-$2.76 & 0.11 & $-$1.86 & 0.15  & $-$0.95 & 0.12\\
G 275-4	    & 5942 & 4.05 & $-$3.42 & 0.09 & $-$2.48 & 0.15  & $-$1.53 & 0.11\\

\tablenotetext{a}{These are the standard deviations based on the agreement of 
the different Fe I lines used.}
\enddata 
\end{deluxetable} 

\clearpage

\singlespace

\begin{deluxetable}{lcrccrccccc}
\tablewidth{0pc}
\tablenum{4}
\tablecaption{New Beryllium and Oxygen Abundances from Boesgaard et
al.~(1999a)}
\tablecolumns{11}
\tablehead{
\colhead{Star}& \colhead{$T_{\rm eff}$ (K)}& \colhead{$\sigma$}& 
\colhead{log g}& \colhead{$\sigma$}&
\colhead{[Fe/H]}& \colhead{$\sigma$}& \colhead{[O/H]}& \colhead{$\sigma$}& 
\colhead{A(Be)}& \colhead{$\sigma$}}
\startdata
HD 19445&  5853 & 40 & 4.41 & 0.23 & $-$2.10 & 0.12 &$-$1.53 &0.10 & $-$0.48 &
 0.06\\
HD 64090&  5500 & 40 & 4.73 & 0.10 & $-$1.77 & 0.18 &$-$1.36 &0.07 & $-$0.09 &
 0.13\\
HD 74000&  6134 & 40 & 4.26 & 0.10 & $-$2.05 & 0.12 &$-$1.56 &0.06 & $-$0.49 &
 0.06\\
HD 76932&  5807 & 40 & 4.00 & 0.12 & $-$0.95 & 0.11 &$-$0.65 &0.08 & +0.77   &
 0.05\\
HD 84937&  6206 & 40 & 3.89 & 0.13 & $-$2.20 & 0.14 &$-$1.49 &0.05 & $-$0.83 &
 0.07\\
HD 94028&  5907 & 64 & 4.44 & 0.17 & $-$1.54 & 0.09 &$-$1.17 &0.10 & +0.39   &
 0.08\\
HD 134169& 5759 & 82 & 3.68 & 0.16 & $-$0.94 & 0.05 &$-$0.66 &0.10 & +0.55   &
 0.10\\
HD 140283& 5692 & 40 & 3.47 & 0.13 & $-$2.56 & 0.12 &$-$1.72 &0.06 & $-$1.18 &
 0.09\\
HD 184499& 5670 & 40 & 4.00 & 0.15 & $-$0.51 & 0.14 &$-$0.41 &0.08 & +1.12   &
 0.09\\
HD 194598& 5911 & 42 & 4.32 & 0.13 & $-$1.25 & 0.16 &$-$1.06 &0.07 & +0.14   & 
0.05\\
HD 201889& 5553 & 41 & 3.74 & 0.24 & $-$0.95 & 0.26 &$-$1.02 &0.08 & +0.58   & 
0.13\\
HD 219617& 5872 & 40 & 4.52 & 0.15 & $-$1.58 & 0.16 &$-$1.27 &0.08 & $-$0.23 &
 0.12\\
BD +37\arcdeg 1458&   5500& 40&	 3.41& 0.26& $-$2.14& 0.17& $-$1.14 & 0.09 &
$-$0.93& 0.08 \\
BD +26\arcdeg 3578&   6158& 64&	 3.94& 0.12& $-$2.32& 0.15& $-$1.69 & 0.06 &
$-$0.90& 0.10 \\
BD +23\arcdeg 3912&   5691& 40&	 3.68& 0.10& $-$1.53& 0.15& $-$1.17 & 0.06 &
+0.02&   0.05 \\
BD +20\arcdeg 3603&   6114& 89&	 4.27& 0.20& $-$2.22& 0.16& $-$1.66 & 0.12 &
$-$0.60& 0.11 \\
BD +17\arcdeg 4708&   5956& 62&	 3.65& 0.26& $-$1.81& 0.10& $-$1.31 & 0.09 &
$-$0.43& 0.08 \\
BD +03\arcdeg 740&    6110& 82&	 3.64& 0.33& $-$2.89& 0.09& $-$2.16 & 0.12 &
$-$1.37& 0.10 \\
BD +02\arcdeg 3375&   5800& 40&	 4.07& 0.41& $-$2.39& 0.17& $-$1.85 & 0.11 &
$-$0.74& 0.15 \\
BD $-$04\arcdeg	3208& 6316& 121& 3.90& 0.42& $-$2.35& 0.20& $-$1.37 & 0.11 &
$-$0.72& 0.15 \\
BD $-$13\arcdeg 3442& 5900& 40&  3.50& 0.44& $-$3.02& 0.16& $-$2.28 & 0.12 &
$-$1.70& 0.21 \\

\enddata
\end{deluxetable}


\clearpage

\singlespace

\begin{deluxetable}{lcrccrccccc}
\tablewidth{0pc}
\tablenum{5}
\tablecaption{Other Beryllium and Oxygen Abundances from Subaru and Keck 
Observations}
\tablecolumns{11}
\tablehead{
\colhead{Star}& \colhead{$T_{\rm eff}$ (K)}& \colhead{$\sigma$}& 
\colhead{log g}& \colhead{$\sigma$}&
\colhead{[Fe/H]}& \colhead{$\sigma$}& \colhead{[O/H]}& \colhead{$\sigma$}& 
\colhead{A(Be)}& \colhead{$\sigma$}}
\startdata
HD 24289\tablenotemark{a} &  5700 & 100 & 3.50 & 0.50 & $-$2.22 & 0.10
&$-$1.67 &0.14 & $-$0.83 & 0.14\\
HD 233511\tablenotemark{a}&  5900 & 40 & 4.20 & 0.06 & $-$1.70 & 0.08 &$-$1.23
&0.14 & $-$0.23 & 0.14\\
HD 132475\tablenotemark{b}&  5765 & 62 & 3.60 & 0.20 & $-$1.50 & 0.15 &$-$0.90
&0.15 & +0.57 & 0.12\\
G 21-22\tablenotemark{b}  &  5916 & 84 & 4.59 & 0.30 & $-$1.02 & 0.25 &$-$0.98
&0.15 & +0.33   & 0.16\\
HD 194598\tablenotemark{a}&  5875 & 40 & 4.2 & 0.06 & $-$1.23 & 0.08 &$-$1.00
&0.14 & +0.12 & 0.14\\
HD 195633\tablenotemark{b}&  5986 & 115 & 3.89 & 0.08 & $-$0.88 & 0.16
&$-$0.67 & 0.15 & +0.66  & 0.11\\
HD 201891\tablenotemark{b}& 5806 & 110 & 4.42 & 0.20 & $-$1.07 & 0.10 &$-$0.86
& 0.15 & +0.62  & 0.11\\

\enddata
\tablenotetext{a}{Keck data in Boesgaard (2007)}
\tablenotetext{b}{Subaru data in Boesgaard \& Novicki (2006)}

\end{deluxetable}


\clearpage
\begin{figure}
\plotone{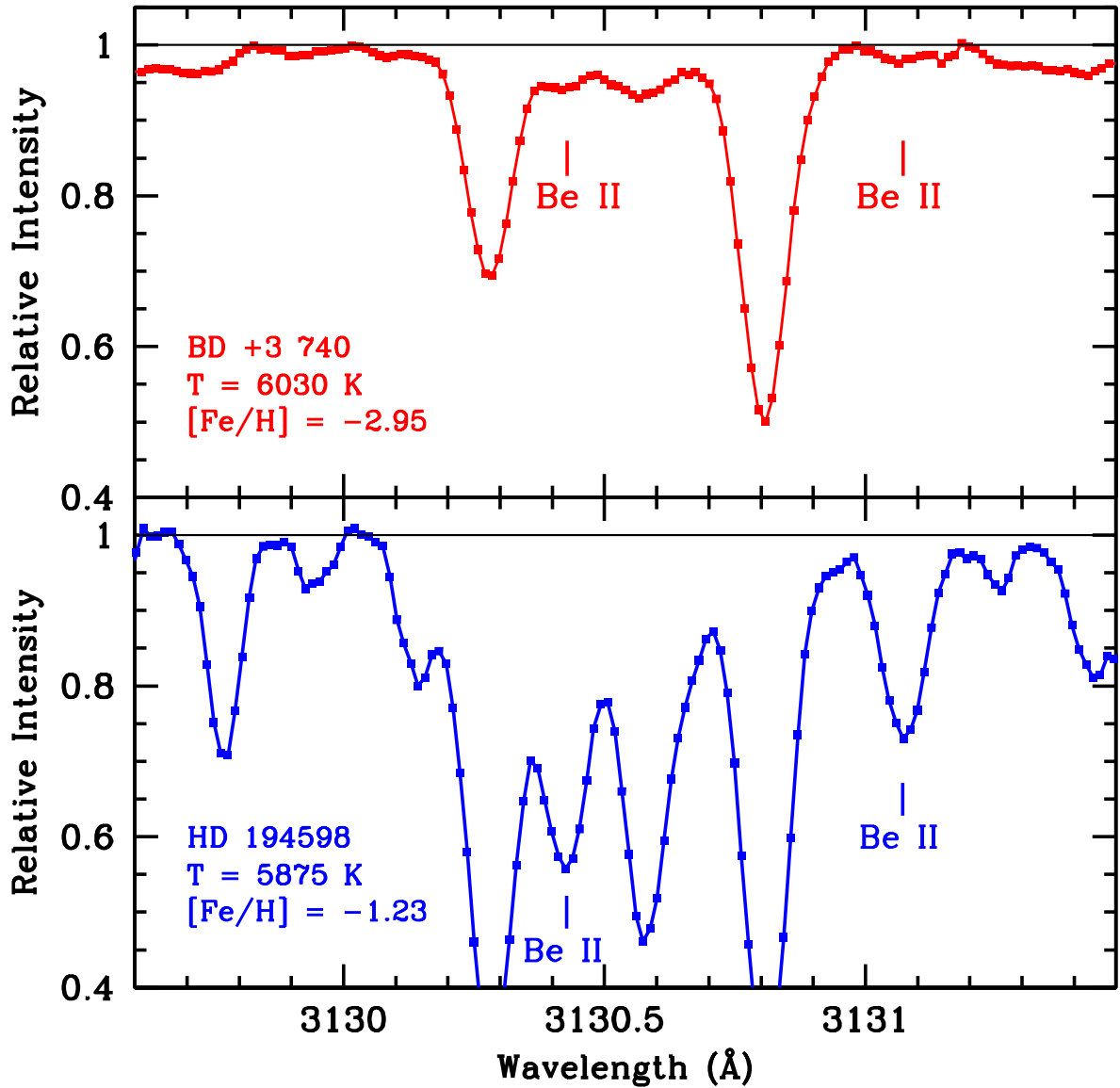}
\caption{The Be region of BD +3 740 ([Fe/H] = $-$2.95) from our sample of very
metal-poor stars compared to HD 194598 ([Fe/H] = $-$1.23) with much higher
metallicity.  Both the Be doublet and the OH feature (at 3130.6 \AA) are much
reduced at low metallicities, necessitating the high signal to noise of our
observations.}
\end{figure}  

\begin{figure}
\plotone{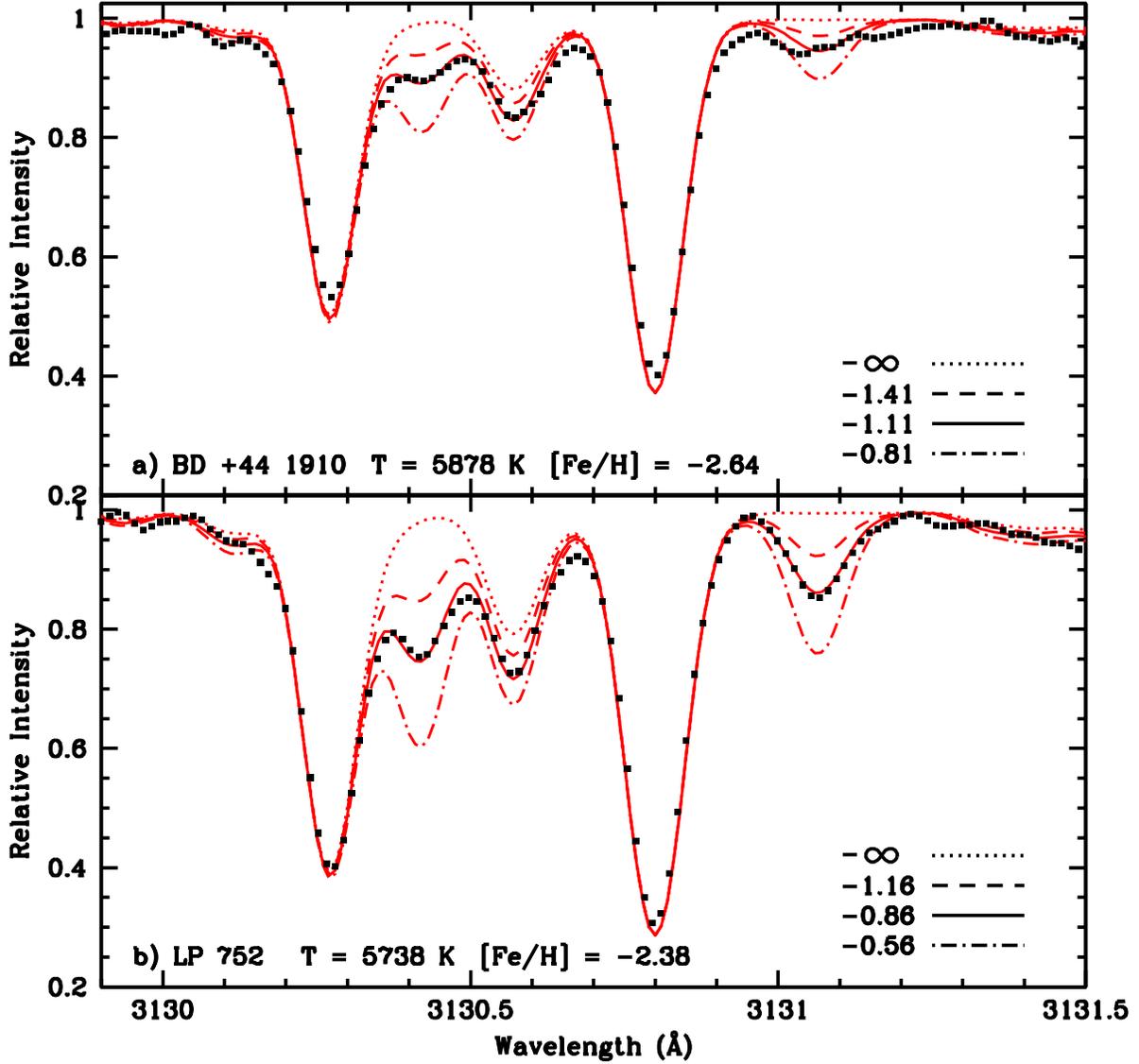}
\caption{Examples of a output from MOOG's ``synth'' driver.  The small squares
are the spectral data and the solid lines are the best fits.  The dotted line
represents a fit with no Be, while the dot-dash line has a factor of 2 more Be
than the best fit and the dashed line has a factor of 2 less.  The OH feature
in the region at 3130.57 \AA\ is used to find the O abundance.  The O
abundance has also been varied by $\pm$0.1 dex in the models.}
\end{figure}

\begin{figure}
\plotone{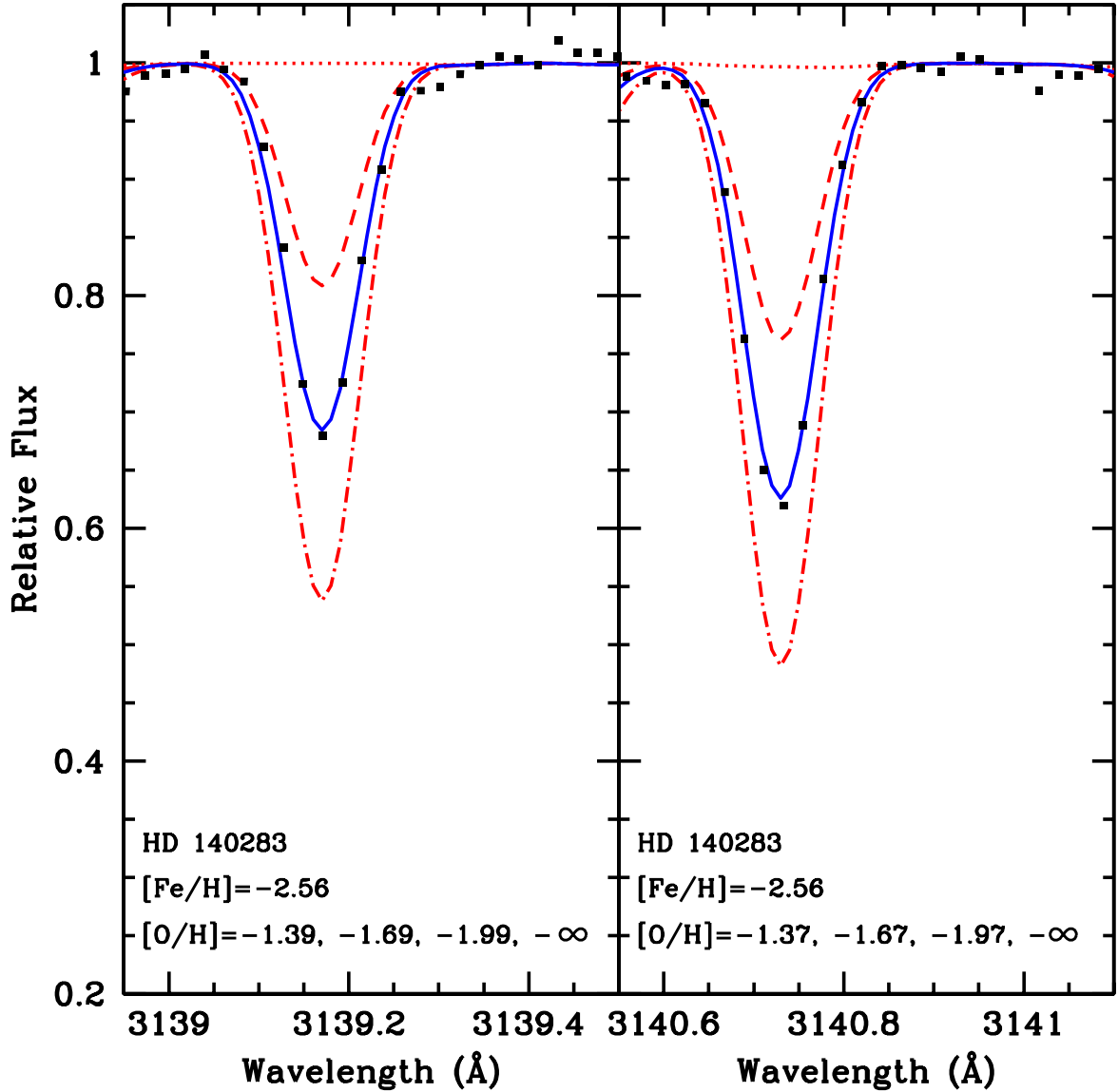}
\caption{An example of the spectrum synthesis for the other two OH regions
that we used to find the O abundances.  The small squares are the spectral
data and the solid lines are the best fits.  The dot-dash line has a factor of
2 more O than the best fit and the dashed line has a factor of 2 less.}
\end{figure}

\begin{figure}
\plotone{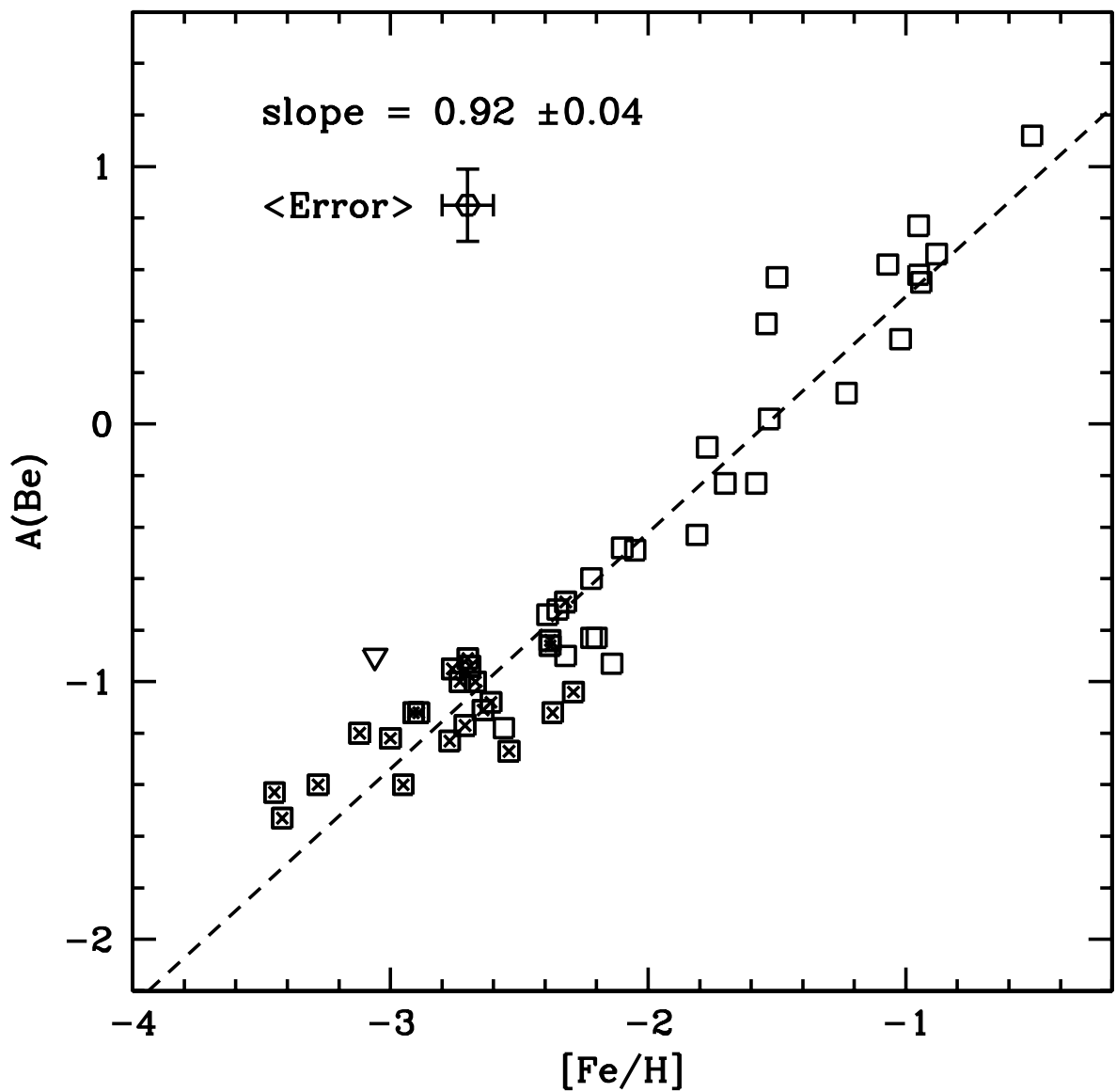}
\caption{Our Be abundances versus [Fe/H] along with a typical error bar.  The
squared crosses represent the new data in this paper while the open squares
are revised values from Tables 4 and 5.  The triangle is the Be upper limit
for LP 831-70, which was not included in the least squares fit.  The results
do not seem to indicate a significant Be plateau at low metallicities, a
possibility discussed in Primas et al. (2000a).  The results can be
represented by a single linear fit over more than 3 orders of magnitude in
[Fe/H] and more than 2 orders of magnitude in A(Be).}
\end{figure}

\begin{figure}
\plotone{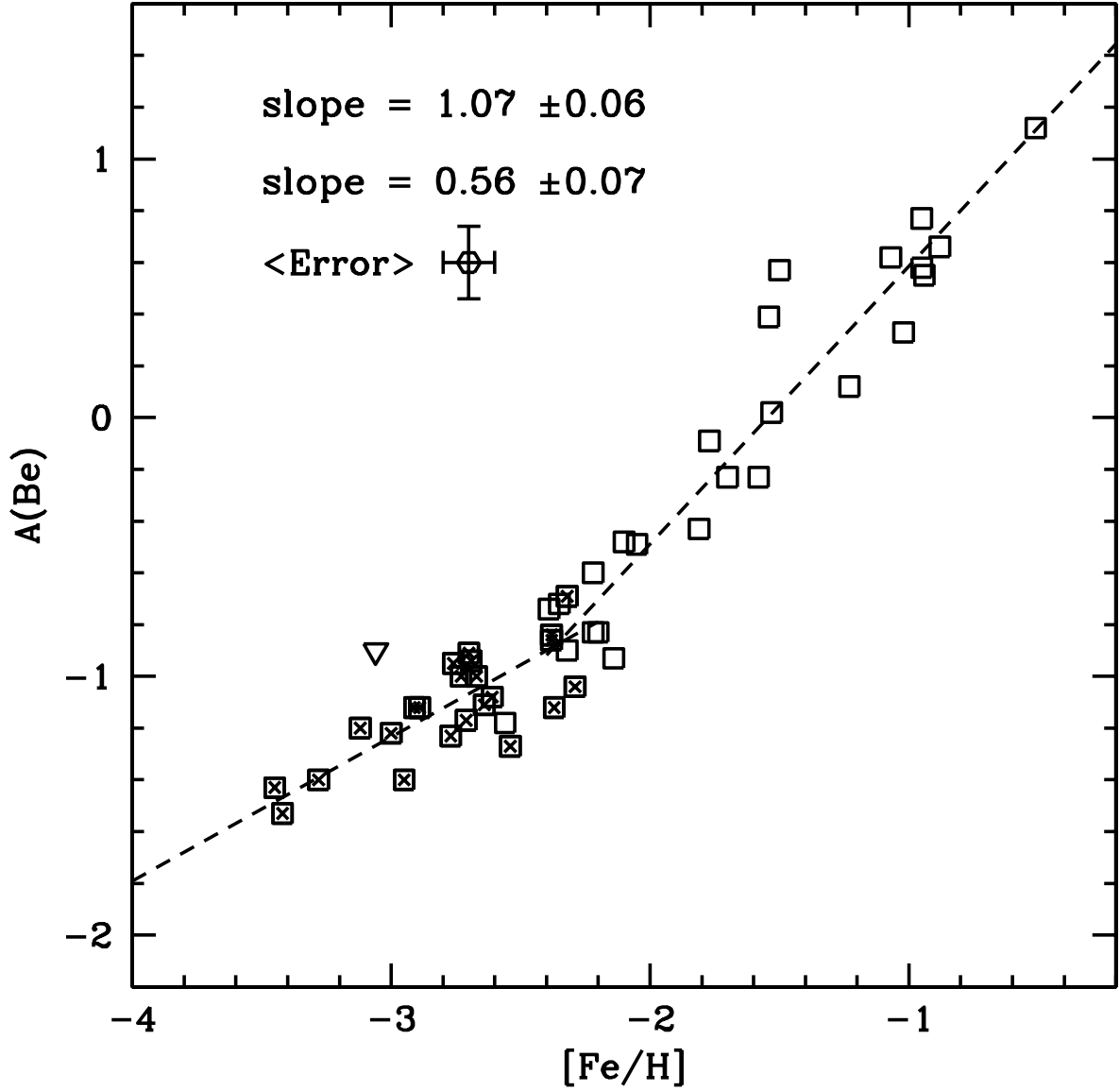}
\caption{This is the same plot as Figure 4, but with a dual linear fit.  This
fit may be more pleasing to the eye, but there is no statistical reason to
prefer it.}
\end{figure}

\begin{figure}
\plotone{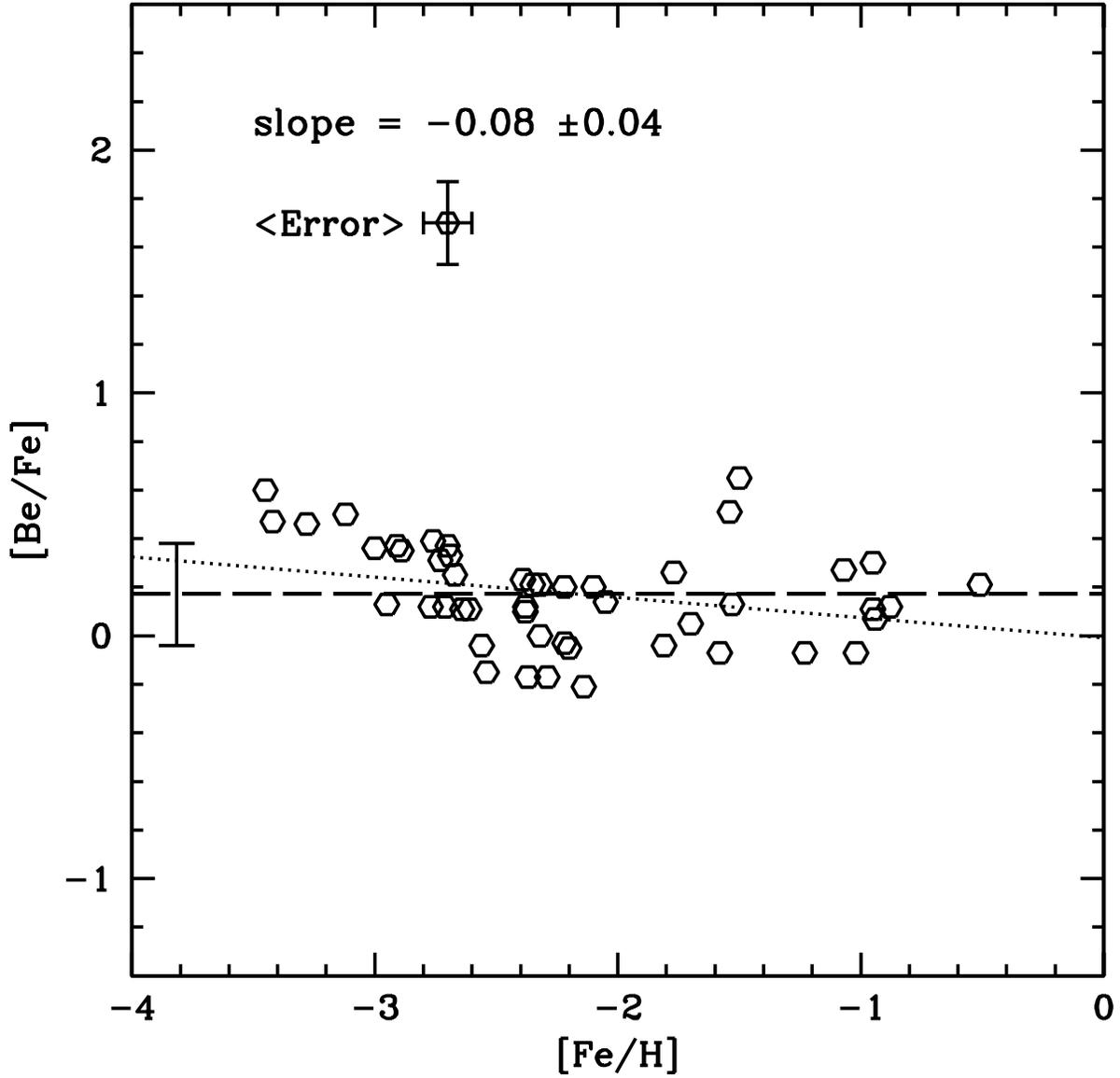}
\caption{This figure shows the trend of Be as normalized to Fe, [Be/Fe],
compared to the Fe abundance.  The dashed line is the mean value of [Be/Fe],
+0.17 $\pm$0.21, and the dotted line is the best linear fit.  The probable
error of the mean value is shown at the left along the dashed mean line.  The
most metal poor stars show an upward trend in [Be/Fe] toward lower [Fe/H],
rather than a plateau.}
\end{figure}

\begin{figure}
\plotone{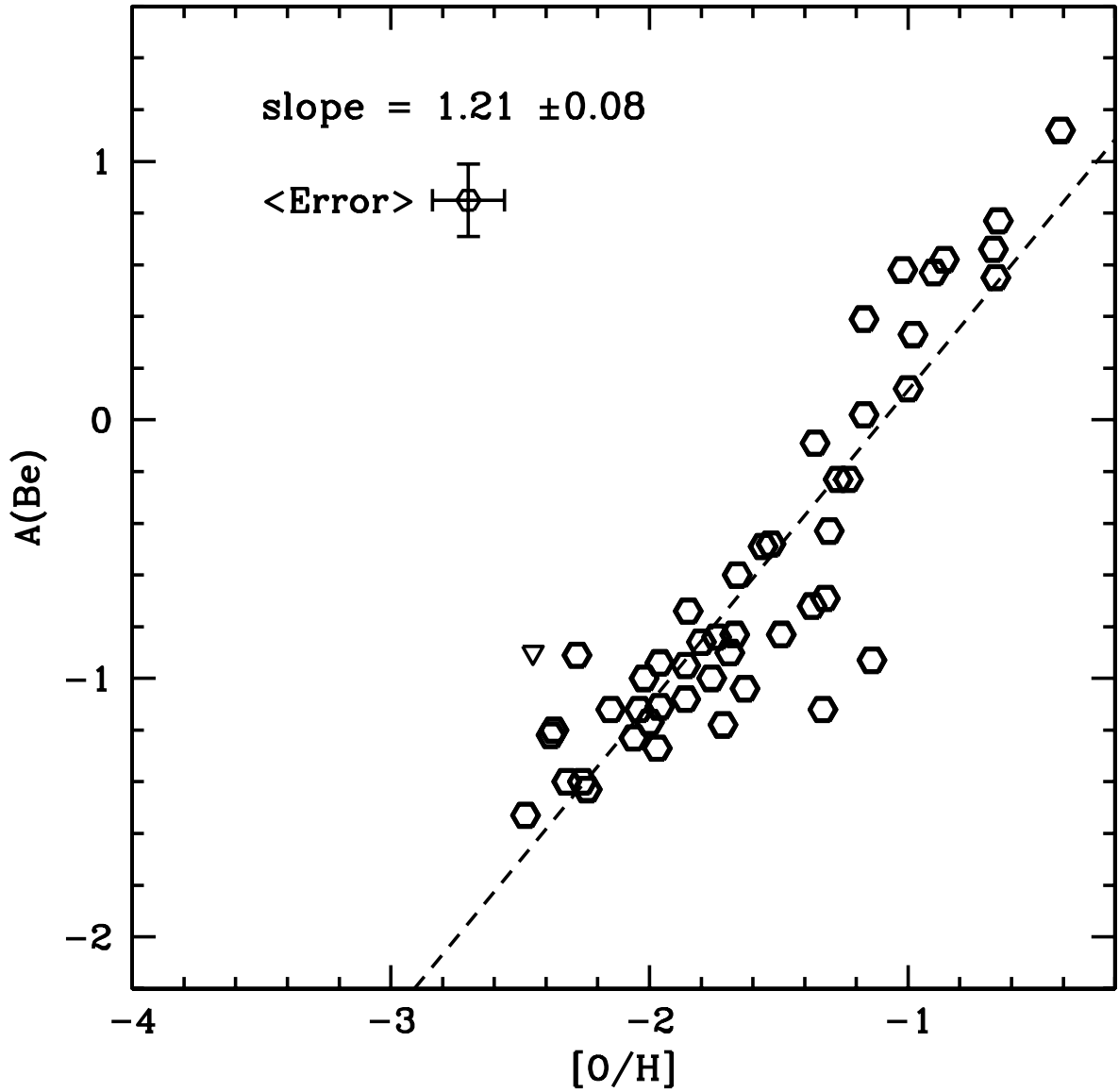}
\caption{The Be abundances compared to our O abundances.  The triangle is the
Be upper limit for LP 831-70, which was not included in the least squares fit.
A linear fit is shown between the two quantities and has a slope of 1.21
$\pm$0.08.  The two stars with high [O/H] and/or low A(Be) are G 108-58 and BD
+37 1458.}
\end{figure}

\begin{figure}
\plotone{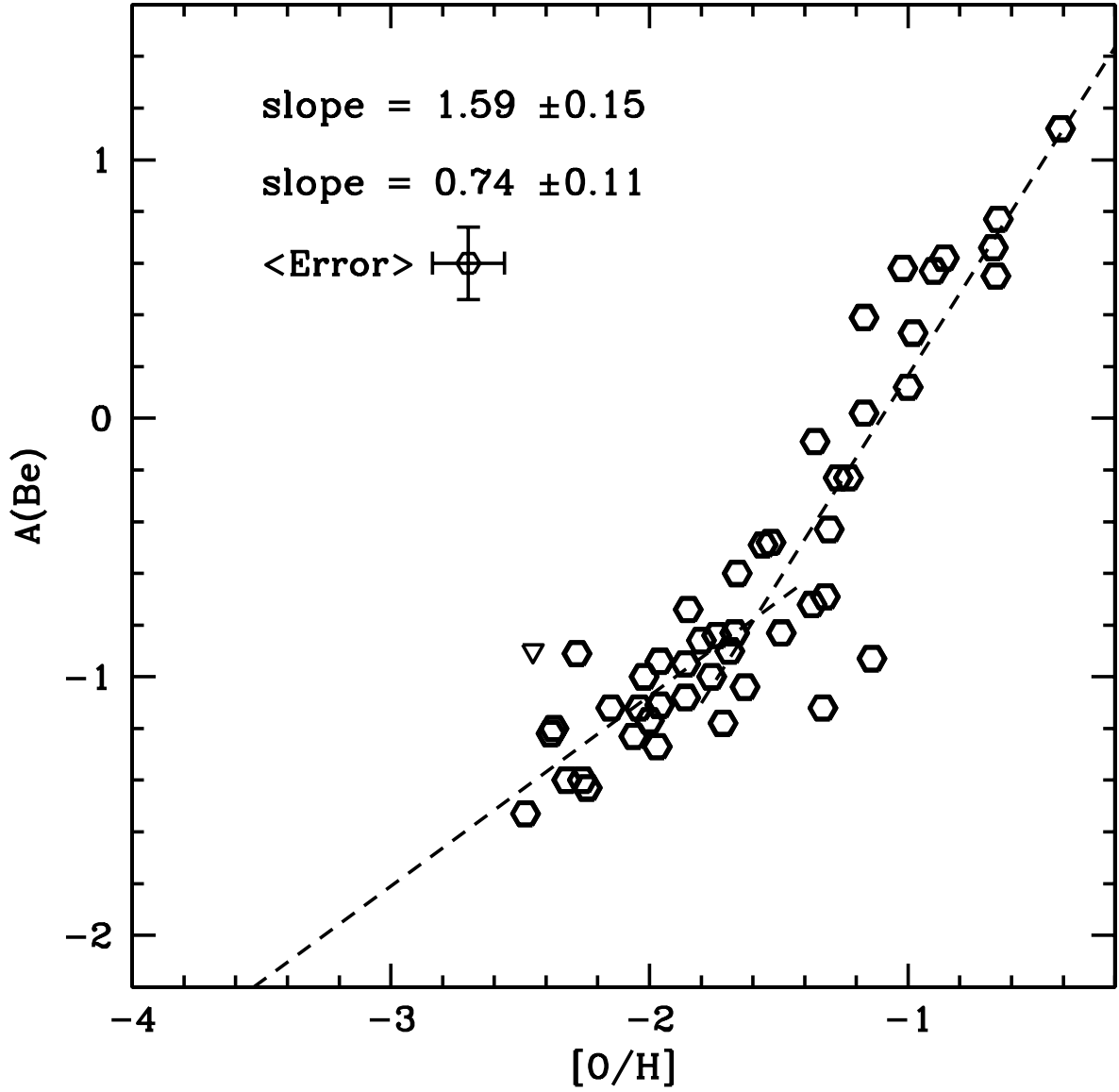}
\caption{The same as Figure 7 except the data are fit with two linear
relations: one for stars with [O/H] $>$ $-$1.8 and the other for stars with
[O/H] $<$ $-$1.4.  The transition seems to occur at [O/H] near $-$1.6,
corresponding to [Fe/H] $\sim-$2.2.}
\end{figure}


\begin{figure}
\plotone{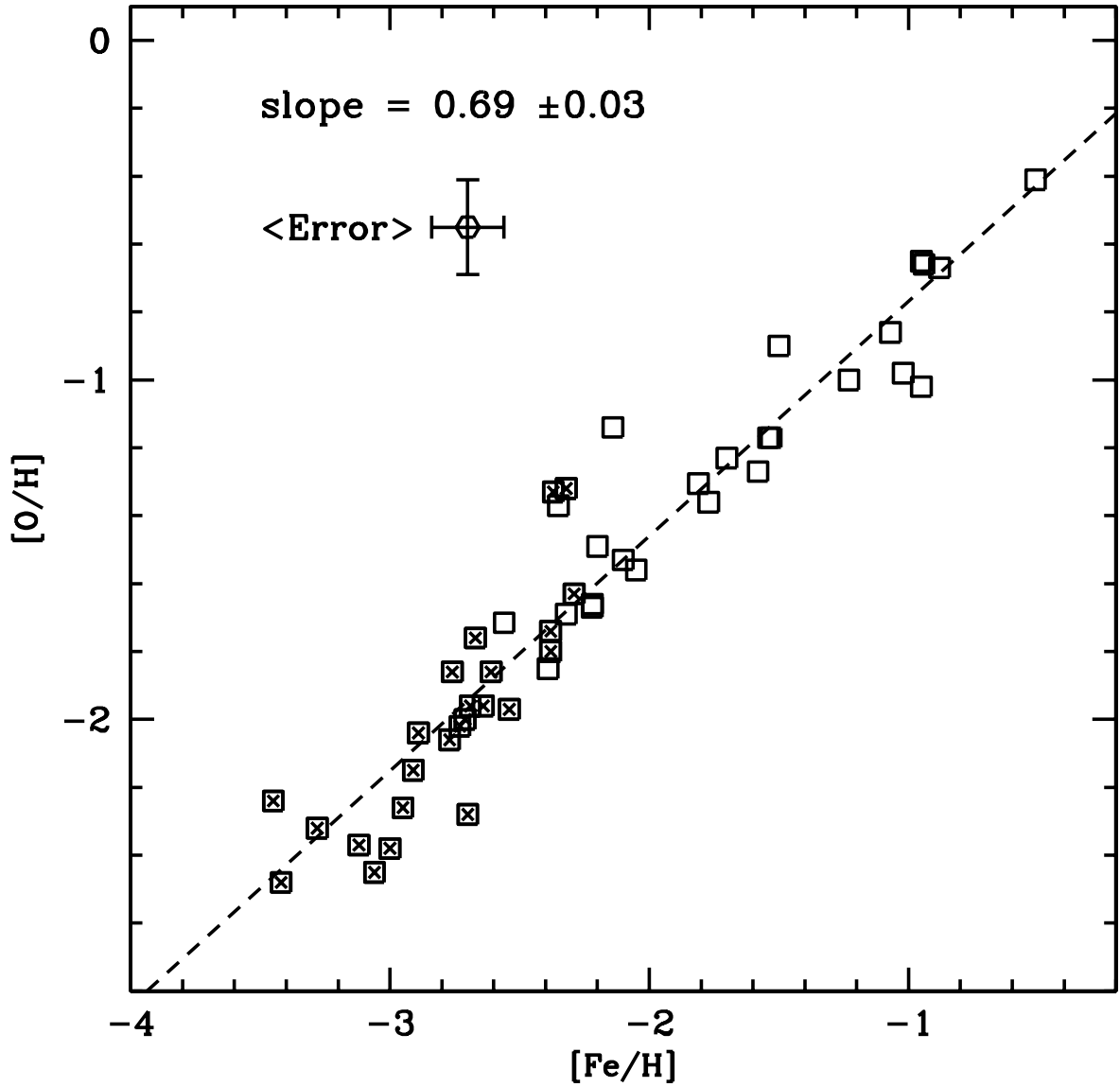}
\caption{The tight linear relationship between O and Fe.  The fit is good over
the 3+ orders of magnitude in Fe and 2+ orders of magnitude in O.  The squared
crosses are the new data from Table 3 and the open squares are the data from
Tables 4 and 5.  The data sets merge well with each other.}
\end{figure}

\begin{figure}
\plotone{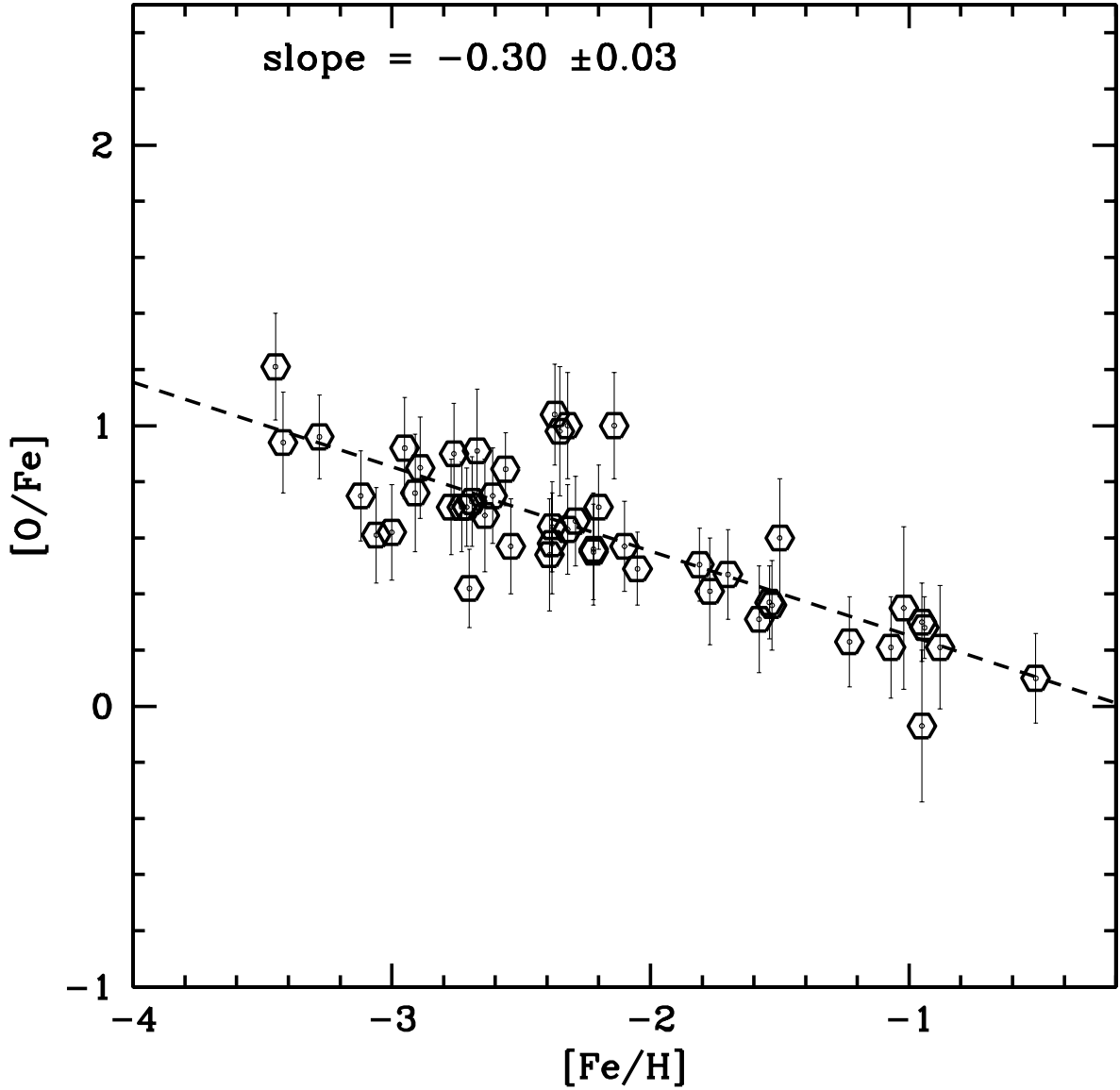}
\caption{This is a standard plot showing the enhancement of O relative to Fe
in low metal stars.  It indicates that the O enhancement diminishes steadily
reaching [O/Fe] = 0.0 at [Fe/H] = 0.0.  In this Figure we show the individual
error bars on [O/Fe].}
\end{figure}

\clearpage
\section{Appendix}
\tightenlines
\singlespace


\begin{deluxetable} {clccccccccccccccc}
\small
\tablewidth{0pc}
\tablenum{6a}
\tablecaption{Measured Equivalent Widths}
\tablehead{
\multicolumn{1}{c}{$\lambda$ ($\rm{\AA}$)} &
\multicolumn{1}{l}{Ele.} &
\multicolumn{1}{c}{1} &
\multicolumn{1}{c}{2} &
\multicolumn{1}{c}{3} &
\multicolumn{1}{c}{4} &
\multicolumn{1}{c}{5} & 
\multicolumn{1}{c}{6} &
\multicolumn{1}{c}{7} & 
\multicolumn{1}{c}{8} &
\multicolumn{1}{c}{9} & 
\multicolumn{1}{c}{10} &
\multicolumn{1}{c}{11} & 
\multicolumn{1}{c}{12} 
%
}
\startdata
3100.304 &Fe I  &...  &...  &...  &41.9 &...  &...  &...  &...  &41.5 &...  &44.8 &...   \\
3100.665 &Fe I  &...  &...  &...  &46.4 &...  &...  &...  &...  &43.3 &...  &...  &...   \\
3116.631 &Fe I  &20.0 &...  &22.2 &16.3 &...  &34.7 &...  &17.6 &15.8 &...  &20.4 &20.1  \\
4494.563 &Fe I  &12.7 &31.8 &8.3  &11.7 &...  &24.1 &15.8 &11.2 &11.3 &20.6 &11.0 &15.5  \\
4528.614 &Fe I  &24.6 &...  &17.4 &20.1 &...  &36.9 &29.0 &18.1 &23.0 &...  &21.0 &26.1  \\
4531.148 &Fe I  &6.4 &16.2  &6.1  &6.1  &19.7 &11.3 &9.6  &7.4  &6.8  &16.1 &5.9  &7.0   \\
4556.126 &Fe I  &...  &...  &...  &...  &6.5  &...  &...  &...  &...  &5.8  &...  &...   \\
4592.651 &Fe I  &...  &...  &...  &...  &12.3 &...  &...  &...  &...  &8.2  &...  &...   \\
4602.941 &Fe I  &4.6  &16.4 &...  &4.2  &20.3 &12.7 &7.9  &6.4  &6.3  &12.7 &5.3  &8.6   \\
4647.434 &Fe I  &...  &...  &...  &...  &6.0  &...  &...  &...  &...  &7.6  &...  &...   \\
4678.846 &Fe I  &...  &...  &...  &...  &7.0  &...  &...  &...  &...  &...  &...  &...   \\
4691.411 &Fe I  &...  &...  &...  &...  &5.7  &...  &...  &...  &...  &...  &...  &...   \\ 
4733.591 &Fe I  &...  &...  &...  &...  &5.3  &...  &...  &...  &...  &...  &...  &...   \\ 
4736.773 &Fe I  &5.3  &...  &...  &...  &...  &...  &...  &...  &...  &...  &...  &...   \\
4871.318 &Fe I  &16.7 &14.3 &...  &...  &14.3 &10.9 &7.9  &...  &4.9  &9.9  &4.9  &6.5   \\
4872.137 &Fe I  &12.9 &28.2 &8.6  &7.6  &27.9 &19.0 &12.8 &8.5  &9.2  &17.4 &9.9  &12.2  \\
4890.750 &Fe I  &16.3 &34.3 &10.9 &14.1 &...  &25.9 &20.4 &11.5 &17.3 &24.1 &14.0 &17.6  \\
4891.490 &Fe I  &25.9 &...  &18.5 &21.4 &...  &38.3 &29.0 &20.7 &26.3 &...  &23.2 &26.8  \\
4918.994 &Fe I  &17.6 &...  &13.4 &13.9 &...  &31.2 &19.2 &12.0 &15.4 &26.7 &14.5 &20.7  \\
4920.503 &Fe I  &32.1 &...  &27.2 &31.4 &...  &46.7 &...  &31.9 &31.3 &...  &29.0 &...   \\
4985.253 &Fe I  &...  &...  &...  &...  &5.7  &...  &...  &...  &...  &...  &...  &...   \\
4985.547 &Fe I  &...  &...  &...  &...  &6.9  &...  &...  &...  &...  &...  &...  &...   \\
4994.130 &Fe I  &...  &...  &...  &...  &15.0 &...  &...  &...  &...  &...  &...  &...   \\
5001.862 &Fe I  &...  &...  &...  &...  &14.1 &...  &...  &...  &...  &...  &...  &...   \\
5006.119 &Fe I  &9.7  &25.8 &7.9  &10.6 &28.9 &20.5 &13.2 &11.4 &9.6  &18.2 &6.7  &13.0  \\
5012.068 &Fe I  &10.2 &23.0 &7.4  &9.7  &30.9 &16.7 &10.9 &7.5  &7.4  &19.2 &6.9  &10.1  \\
5022.236 &Fe I  &...  &...  &...  &...  &6.3  &...  &...  &...  &...  &...  &...  &...   \\
5049.819 &Fe I  &...  &...  &...  &...  &22.4 &...  &...  &...  &...  &...  &...  &...   \\
5051.635 &Fe I  &...  &...  &...  &...  &24.2 &...  &...  &...  &...  &...  &...  &...   \\
5068.766 &Fe I  &...  &...  &...  &...  &10.5 &...  &...  &...  &...  &...  &...  &...   \\
5079.224 &Fe I  &...  &...  &...  &...  &8.2  &...  &...  &...  &...  &...  &...  &...   \\
5079.740 &Fe I  &...  &...  &...  &...  &8.2  &...  &...  &...  &...  &...  &...  &...   \\
5083.339 &Fe I  &...  &...  &...  &...  &15.3 &...  &...  &...  &...  &...  &...  &...   \\
5098.697 &Fe I  &...  &...  &...  &...  &11.4 &...  &...  &...  &...  &...  &...  &...   \\
5123.720 &Fe I  &3.2  &7.9  &...  &...  &11.9 &7.1  &4.4  &...  &...  &...  &...  &...   \\
5150.840 &Fe I  &...  &...  &...  &...  &11.4 &...  &...  &...  &...  &...  &...  &...   \\
5171.596 &Fe I  &13.9 &33.1 &8.9  &17.7 &...  &34.0 &16.6 &12.6 &10.8 &24.2 &13.0 &17.4  \\
5192.344 &Fe I  &10.1 &27.5 &8.9  &11.4 &30.1 &20.6 &12.7 &8.6  &11.6 &19.1 &8.9  &13.7  \\
5194.942 &Fe I  &...  &...  &...  &...  &...  &...  &...  &...  &...  &...  &...  &...   \\
5198.711 &Fe I  &...  &...  &...  &...  &8.0  &...  &...  &...  &...  &...  &...  &...   \\
5202.336 &Fe I  &...  &...  &...  &...  &13.1 &...  &...  &...  &...  &...  &...  &...   \\
5215.182 &Fe I  &...  &...  &...  &...  &7.4  &...  &...  &...  &...  &...  &...  &...   \\
5216.274 &Fe I  &...  &...  &...  &...  &20.1 &...  &...  &...  &...  &...  &...  &...   \\
5217.390 &Fe I  &...  &...  &...  &...  &6.2  &...  &...  &...  &...  &...  &...  &...   \\
5227.190 &Fe I  &32.5 &...  &22.8 &30.7 &...  &50.2 &...  &28.0 &25.6 &...  &30.0 &36.6  \\
5232.940 &Fe I  &23.0 &...  &15.2 &20.9 &...  &37.1 &27.9 &19.2 &18.6 &35.4 &19.4 &25.8  \\
5242.491 &Fe I  &...  &...  &...  &...  &5.2  &...  &...  &...  &...  &...  &...  &...   \\
5250.646 &Fe I  &...  &...  &...  &...  &6.8  &...  &...  &...  &...  &...  &...  &...   \\
5263.305 &Fe I  &...  &...  &...  &...  &8.5  &...  &...  &...  &...  &...  &...  &...   \\
5269.537 &Fe I  &...  &...  &51.2 &57.6 &...  &77.2 &...  &53.7 &54.2 &...  &59.0 &...   \\
5328.039 &Fe I  &...  &...  &43.1 &49.0 &...  &68.5 &...  &45.0 &44.7 &...  &50.4 &...   \\
5332.900 &Fe I  &...  &...  &...  &...  &5.1  &...  &...  &...  &...  &...  &...  &...   \\
5339.930 &Fe I  &...  &...  &...  &...  &14.9 &...  &...  &...  &...  &...  &...  &...   \\
5341.024 &Fe I  &...  &27.7 &11.9 &9.2  &28.0 &18.7 &...  &5.9  &10.2 &19.7 &10.9 &11.6  \\
5393.167 &Fe I  &...  &...  &...  &...  &11.3 &...  &...  &...  &...  &...  &...  &...   \\
5397.128 &Fe I  &25.4 &...  &19.1 &23.5 &...  &41.5 &32.7 &22.7 &20.5 &     &22.4 &29.4  \\
5405.775 &Fe I  &28.8 &...  &21.7 &31.2 &...  &45.7 &35.2 &24.5 &23.1 &...  &26.1 &31.6  \\
5429.696 &Fe I  &29.8 &...  &22.3 &26.8 &...  &46.9 &...  &25.3 &26.9 &...  &26.1 &32.8  \\
5434.524 &Fe I  &17.4 &...  &12.4 &22.  &36.5 &30.8 &24.2 &12.3 &12.5 &30.8 &15.7 &20.0  \\
5455.609 &Fe I  &19.1 &...  &15.2 &18.6 &...  &34.6 &28.6 &17.1 &16.1 &35.2 &16.3 &25.4  \\
5497.516 &Fe I  &...  &...  &...  &...  &...  &...  &...  &...  &...  &...  &...  &...   \\
5501.465 &Fe I  &...  &...  &...  &...  &...  &...  &...  &...  &...  &...  &...  &...   \\
5569.618 &Fe I  &...  &...  &...  &...  &12.0 &...  &...  &...  &...  &...  &...  &...   \\
5572.841 &Fe I  &5.8  &20.6 &5.1  &6.2  &20.7 &15.3 &11.8 &...  &7.0  &13.8 &6.6  &10.7  \\
5576.090 &Fe I  &...  &...  &...  &...  &8.3  &...  &...  &...  &...  &...  &...  &...   \\
5586.756 &Fe I  &...  &29.9 &8.6  &12.2 &30.9 &21.5 &17.6 &10.8 &9.8  &17.9 &12.0 &14.7  \\
5658.816 &Fe I  &...  &...  &...  &...  &8.3  &...  &...  &...  &...  &...  &...  &...   \\

\\
4508.289 &Fe II &6.1  &17.2 &5.3  &7.7  &9.6  &10.8 &9.5  &6.3  &8.6  &12.1 &6.5  &10.5  \\
4515.339 &Fe II &5.5  &14.8 &5.7  &5.5  &8.2  &10.4 &8.4  &4.9  &5.4  &9.5  &5.9  &9.6   \\
4522.634 &Fe II &10.3 &24.6 &7.1  &11.2 &13.9 &14.9 &13.3 &13.4 &10.1 &16.7 &9.5  &15.1  \\
4576.339 &Fe II &...  &6.5  &...  &...  &...  &...  &...  &...  &...  &...  &...  &5.0   \\
4583.837 &Fe II &17.6 &...  &9.6  &21.4 &25.1 &27.8 &23.4 &21.0 &18.5 &28.4 &20.7 &26.8  \\
4629.339 &Fe II &7.4  &18.6 &6.4  &6.9  &9.1  &10.8 &8.9  &7.5  &7.0  &12.0 &6.6  &9.8   \\
5197.576 &Fe II &4.7  &13.5 &...  &4.2  &6.1  &6.7  &6.5  &4.8  &5.7  &6.6  &4.8  &6.1   \\
5234.630 &Fe II &3.6  &15.3 &...  &5.8  &7.7  &9.2  &7.5  &4.0  &5.8  &9.2  &5.3  &8.7   \\
5276.002 &Fe II &4.5  &18.6 &...  &8.3  &8.2  &10.7 &8.3  &8.5  &7.0  &11.9 &7.0  &9.2   \\
\\
4518.023 &Ti I  &...  &5.1  &...  &...  &...  &5.0  &...  &...  &...  &...  &...  &...   \\
4527.305 &Ti I  &...  &...  &...  &...  &...  &...  &...  &...  &...  &...  &...  &...   \\
4533.239 &Ti I  &9.7  &19.5 &5.3  &8.8  &16.3 &13.0 &10.4 &8.4  &9.3  &16.6 &9.3  &11.9  \\
4534.778 &Ti I  &...  &16.5 &...  &5.9  &12.0 &13.6 &6.5  &6.4  &4.5  &9.7  &10.4 &10.6  \\
4535.570 &Ti I  &...  &12.6 &...  &4.7  &9.0  &8.7  &4.9  &5.1  &...  &5.3  &7.7  &9.5   \\
4681.908 &Ti I  &...  &5.9  &...  &...  &5.6  &...  &3.6  &...  &...  &5.9  &...  &...   \\
4981.732 &Ti I  &12.9 &23.8 &5.6  &13.6 &18.0 &17.9 &12.4 &8.5  &8.5  &16.9 &11.8 &14.2  \\
4991.067 &Ti I  &13.2 &23.5 &10.9 &7.1  &17.9 &13.8 &15.6 &5.7  &11.2 &20.1 &13.9 &11.9  \\
4999.504 &Ti I  &7.8  &15.9 &...  &...  &12.1 &13.1 &8.6  &5.9  &5.7  &11.0 &7.2  &9.9   \\
5016.162 &Ti I  &...  &3.6  &...  &...  &...  &...  &...  &...  &...  &...  &...  &...   \\
5020.028 &Ti I  &...  &7.5  &...  &...  &5.3  &...  &...  &...  &...  &5.2  &...  &...   \\
5022.871 &Ti I  &...  &...  &...  &...  &...  &...  &...  &...  &...  &...  &...  &...   \\
5035.907 &Ti I  &...  &...  &...  &...  &...  &...  &...  &...  &...  &...  &...  &...   \\
5036.468 &Ti I  &...  &5.9  &...  &...  &...  &...  &...  &...  &...  &...  &...  &...   \\
5039.959 &Ti I  &...  &6.4  &...  &...  &...  &...  &...  &...  &...  &...  &...  &...   \\
5064.654 &Ti I  &...  &5.8  &...  &...  &...  &...  &3.2  &...  &...  &5.4  &...  &...   \\
\\
4501.272 &Ti II &...  &...  &17.5 &40.4 &...  &47.9 &...  &35.6 &34.6 &...  &43.1 &...   \\
4563.761 &Ti II &29.9 &...  &14.0 &33.8 &...  &40.7 &...  &27.5 &27.9 &...  &33.9 &...   \\
4571.968 &Ti II &...  &...  &17.2 &40.8 &...  &45.7 &...  &34.8 &34.9 &...  &41.0 &...   \\
4589.958 &Ti II &6.8  &17.4 &...  &6.6  &7.1  &11.8 &8.5  &7.6  &6.3  &14.6 &6.6  &9.7   \\
4657.203 &Ti II &...  &5.1  &...  &...  &...  &...  &...  &...  &...  &...  &...  &...   \\
4779.985 &Ti II &...  &10.0 &...  &...  &...  &5.5  &...  &...  &...  &6.0  &...  &...   \\
5129.152 &Ti II &...  &9.6  &...  &...  &7.4  &6.3  &5.0  &...  &...  &9.2  &...  &9.3   \\
5154.070 &Ti II &...  &8.7  &...  &...  &6.1  &...  &5.6  &...  &...  &5.6  &...  &...   \\
5188.680 &Ti II &7.9  &27.6 &...  &13.2 &15.6 &15.7 &16.5 &9.0  &12.0 &22.2 &14.8 &13.3  \\
5226.543 &Ti II &7.3  &21.3 &...  &12.9 &11.2 &12.6 &11.8 &6.5  &8.6  &16.2 &9.2  &12.3  \\
5336.781 &Ti II &...  &12.5 &...  &...  &7.5  &8.3  &5.6  &...  &4.2  &6.8  &...  &7.6   \\
5381.018 &Ti II &...  &9.7  &...  &...  &...  &...  &...  &...  &...  &...  &...  &...    \\
\enddata
\end{deluxetable}

\clearpage
\tightenlines
\singlespace


\begin{deluxetable} {clcccccccccccccc}
\small
\tablewidth{0pc}
\tablenum{6b}
\tablecaption{Measured Equivalent Widths}
\tablehead{
\multicolumn{1}{c}{$\lambda$ ($\rm{\AA}$)} &
\multicolumn{1}{l}{Ele.} &
\multicolumn{1}{c}{13} &
\multicolumn{1}{c}{14} &
\multicolumn{1}{c}{15} &
\multicolumn{1}{c}{16} &
\multicolumn{1}{c}{17} & 
\multicolumn{1}{c}{18} &
\multicolumn{1}{c}{19} & 
\multicolumn{1}{c}{20} &
\multicolumn{1}{c}{21} & 
\multicolumn{1}{c}{22} &
\multicolumn{1}{c}{23}  
}
%
\startdata
3100.304 &Fe I  &...  &...  &...  &...  &46.4 &...  &...  &...  &...  &...  &...      \\
3100.665 &Fe I  &...  &...  &...  &...  &...  &...  &...  &...  &...  &...  &...      \\
3116.631 &Fe I  &20.1 &...  &11.6 &...  &14.7 &16.6 &...  &26.5 &29.9 &15.1 &...      \\
4494.563 &Fe I  &15.5 &26.8 &4.9  &24.1 &12.6 &9.9  &...  &15.3 &18.8 &9.1  &5.6      \\
4528.614 &Fe I  &26.1 &...  &8.2  &...  &23.5 &21.2 &...  &27.8 &31.6 &17.1 &8.0      \\
4531.148 &Fe I  &7.0  &16.2 &...  &15.2 &8.4  &7.7  &25.1 &8.6  &11.3 &4.4  &...      \\
4556.126 &Fe I  &...  &7.6  &...  &...  &...  &...  &...  &...  &...  &...  &...      \\
4592.651 &Fe I  &...  &10.6 &...  &...  &...  &...  &...  &...  &...  &...  &...      \\
4602.941 &Fe I  &8.6  &15.4 &...  &12.6 &5.0  &5.1  &23.3 &9.2  &10.3 &3.1  &...      \\
4647.434 &Fe I  &...  &6.1  &...  &...  &...  &...  &...  &...  &...  &...  &...      \\
4678.846 &Fe I  &...  &5.3  &...  &...  &...  &...  &...  &...  &...  &...  &...      \\
4691.411 &Fe I  &...  &...  &...  &...  &...  &...  &...  &...  &...  &...  &...      \\ 
4733.591 &Fe I  &...  &...  &...  &...  &...  &...  &...  &...  &...  &...  &...      \\ 
4736.773 &Fe I  &6.5  &13.5 &...  &12.2 &...  &...  &19.9 &7.2  &9.1  &4.6  &...      \\
4871.318 &Fe I  &17.6 &30.1 &6.4  &30.3 &16.8 &13.1 &...  &20.7 &22.0 &17.0 &...      \\
4872.137 &Fe I  &12.2 &21.8 &...  &22.7 &12.4 &8.7  &33.4 &16.0 &17.8 &12.8 &...      \\
4890.750 &Fe I  &17.6 &31.7 &5.0  &28.7 &15.5 &13.4 &...  &19.7 &22.4 &10.9 &5.6      \\
4891.490 &Fe I  &26.8 &...  &10.2 &...  &24.9 &23.0 &...  &30.6 &35.0 &18.6 &8.3      \\
4918.994 &Fe I  &20.7 &34.8 &13.0 &26.8 &...  &12.8 &...  &20.6 &24.4 &12.6 &6.6      \\
4920.503 &Fe I  &...  &...  &18.0 &...  &49.4 &28.2 &...  &35.9 &44.3 &24.8 &14.7     \\
4985.253 &Fe I  &...  &...  &...  &...  &...  &...  &...  &...  &...  &...  &...      \\
4985.547 &Fe I  &...  &...  &...  &...  &...  &...  &...  &...  &...  &...  &...      \\
4994.130 &Fe I  &...  &13.9 &...  &...  &...  &...  &...  &...  &...  &...  &...      \\
5001.862 &Fe I  &...  &11.8 &...  &...  &...  &...  &...  &...  &...  &...  &...      \\
5006.119 &Fe I  &13.0 &24.8 &...  &21.6 &10.6 &9.8  &35.1 &11.9 &16.7 &9.0  &5.2      \\
5012.068 &Fe I  &10.1 &24.5 &...  &20.0 &11.0 &9.5  &32.0 &12.4 &18.3 &7.1  &3.6      \\
5022.236 &Fe I  &...  &6.5  &...  &...  &...  &...  &...  &...  &...  &...  &...      \\
5049.819 &Fe I  &...  &16.3 &...  &...  &...  &...  &...  &...  &...  &...  &...      \\
5051.635 &Fe I  &...  &15.4 &...  &...  &...  &...  &...  &...  &...  &...  &...      \\
5068.766 &Fe I  &...  &8.4  &...  &...  &...  &...  &...  &...  &...  &...  &...      \\
5079.224 &Fe I  &...  &5.4  &...  &...  &...  &...  &...  &...  &...  &...  &...      \\
5079.740 &Fe I  &...  &6.0  &...  &...  &...  &...  &...  &...  &...  &...  &...      \\
5083.339 &Fe I  &...  &5.4  &...  &...  &...  &...  &...  &...  &...  &...  &...      \\
5098.697 &Fe I  &...  &...  &...  &...  &...  &...  &...  &...  &...  &...  &...      \\
5123.720 &Fe I  &...  &...  &...  &5.8  &...  &...  &13.1 &...  &6.5  &...  &...      \\
5150.840 &Fe I  &...  &7.5  &...  &...  &...  &...  &...  &...  &...  &...  &...      \\
5171.596 &Fe I  &17.4 &31.6 &5.5  &27.2 &13.3 &12.5 &...  &17.5 &20.6 &8.0  &6.8      \\
5192.344 &Fe I  &13.7 &23.8 &...  &23.3 &11.0 &8.6  &33.2 &13.3 &18.7 &6.4  &6.8      \\
5194.942 &Fe I  &...  &17.8 &...  &...  &...  &...  &...  &...  &...  &...  &...      \\
5198.711 &Fe I  &...  &5.3  &...  &...  &...  &...  &...  &...  &...  &...  &...      \\
5202.336 &Fe I  &...  &10.3 &...  &...  &...  &...  &...  &...  &...  &...  &...      \\
5215.182 &Fe I  &...  &6.4  &...  &...  &...  &...  &...  &...  &...  &...  &...      \\
5216.274 &Fe I  &...  &13.0 &...  &...  &...  &...  &...  &...  &...  &...  &...      \\
5217.390 &Fe I  &...  &6.5  &...  &...  &...  &...  &...  &...  &...  &...  &...      \\
5227.190 &Fe I  &36.6 &...  &10.9 &...  &30.2 &31.0 &...  &35.0 &42.5 &23.1 &15.5      \\
5232.940 &Fe I  &25.8 &...  &7.8  &...  &22.4 &20.1 &...  &28.8 &34.7 &16.0 &8.1      \\
5242.491 &Fe I  &...  &5.6  &...  &...  &...  &...  &...  &...  &...  &...  &...      \\
5250.646 &Fe I  &...  &5.5  &...  &...  &...  &...  &...  &...  &...  &...  &...      \\
5263.305 &Fe I  &...  &6.9  &...  &...  &...  &...  &...  &...  &...  &...  &...      \\
5269.537 &Fe I  &...  &...  &30.2 &...  &58.3 &...  &...  &66.4 &73.9 &...  &36.4      \\
5328.039 &Fe I  &...  &...  &23.5 &...  &51.3 &...  &...  &56.8 &61.7 &...  &28.9      \\
5332.900 &Fe I  &...  &...  &...  &...  &...  &...  &...  &...  &...  &...  &...      \\
5339.930 &Fe I  &...  &11.5 &...  &...  &...  &...  &...  &...  &...  &...  &...      \\
5341.024 &Fe I  &11.6 &23.4 &...  &23.1 &...  &...  &34.3 &...  &15.4 &...  &...      \\
5393.167 &Fe I  &...  &10.8 &...  &...  &...  &...  &...  &...  &...  &...  &...      \\
5397.128 &Fe I  &29.4 &...  &10.1 &...  &27.6 &24.6 &...  &31.3 &36.4 &17.4 &10.8      \\
5405.775 &Fe I  &31.6 &...  &11.5 &...  &27.2 &27.1 &...  &32.2 &40.6 &19.7 &13.1      \\
5429.696 &Fe I  &32.8 &...  &8.8  &...  &29.3 &28.4 &...  &35.9 &42.4 &20.5 &12.8      \\
5434.524 &Fe I  &20.0 &...  &5.8  &33.5 &18.5 &16.3 &...  &21.3 &26.4 &12.0 &8.0      \\
5455.609 &Fe I  &25.4 &...  &5.9  &36.7 &22.3 &19.2 &...  &23.8 &31.8 &13.2 &10.1      \\
5497.516 &Fe I  &...  &13.1 &...  &...  &...  &...  &...  &...  &...  &...  &...      \\
5501.465 &Fe I  &...  &10.0 &...  &...  &...  &...  &...  &...  &...  &...  &...      \\
5569.618 &Fe I  &...  &12.1 &...  &...  &...  &...  &...  &...  &...  &...  &...      \\
5572.841 &Fe I  &10.7 &17.8 &...  &16.0 &9.0  &6.3  &26.6 &9.0  &14.3 &...  &...      \\
5576.090 &Fe I  &...  &...  &...  &...  &...  &...  &...  &...  &...  &...  &...      \\
5586.756 &Fe I  &14.7 &24.4 &4.6  &23.6 &11.3 &9.5  &33.7 &15.7 &16.2 &9.5  &...      \\
5658.816 &Fe I  &...  &6.6  &...  &...  &...  &...  &...  &...  &...  &...  &...      \\
		 								  
\\		 								  
4508.289 &Fe II &10.5 &13.8 &3.8  &11.7 &8.9  &5.4  &26.3 &9.5  &14.3 &5.8  &...       \\
4515.339 &Fe II &9.6  &9.4  &...  &9.1  &...  &3.2  &20.3 &7.9  &10.4 &7.0  &...       \\
4522.634 &Fe II &15.1 &15.7 &7.1  &16.5 &9.3  &6.2  &...  &12.2 &17.9 &9.5  &...       \\
4576.339 &Fe II &5.0  &...  &...  &5.4  &...  &...  &10.1 &...  &...  &...  &...       \\
4583.837 &Fe II &26.8 &28.4 &5.8  &31.0 &18.8 &13.0 &...  &21.0 &29.3 &16.0 &8.5       \\
4629.339 &Fe II &9.8  &11.7 &...  &13.6 &6.9  &4.9  &27.0 &10.5 &12.0 &6.1  &...       \\
5197.576 &Fe II &6.1  &6.7  &...  &10.7 &...  &5.2  &18.2 &8.2  &9.7  &3.5  &...       \\
5234.630 &Fe II &8.7  &8.0  &3.5  &12.1 &...  &...  &22.7 &8.4  &12.2 &5.2  &...       \\
5276.002 &Fe II &9.2  &9.3  &...  &20.2 &...  &5.2  &28.2 &10.6 &15.0 &6.2  &...       \\
\\		 		 					         	  
4518.023 &Ti I  &...  &...  &...  &5.3  &...  &...  &7.0  &...  &...  &...  &...       \\
4527.305 &Ti I  &...  &...  &...  &...  &...  &...  &4.9  &...  &...  &...  &...       \\
4533.239 &Ti I  &...  &...  &4.6  &...  &...  &...  &...  &...  &...  &...  &7.0       \\
4534.778 &Ti I  &11.9 &15.3 &...  &13.9 &9.5  &7.0  &25.9 &...  &12.4 &7.0  &5.5       \\
4535.570 &Ti I  &9.5  &6.2  &...  &5.6  &...  &3.7  &17.2 &7.7  &6.7  &4.1  &...       \\
4681.908 &Ti I  &...  &...  &...  &...  &...  &...  &8.1  &...  &...  &...  &...       \\
4981.732 &Ti I  &14.2 &16.0 &4.3  &14.0 &9.4  &8.4  &...  &10.6 &14.5 &11.1 &...       \\
4991.067 &Ti I  &11.9 &20.9 &4.7  &19.1 &10.5 &9.8  &24.2 &13.3 &16.8 &...  &5.7       \\
4999.504 &Ti I  &9.9  &10.6 &...  &12.6 &...  &4.2  &21.7 &7.7  &11.5 &4.6  &...       \\
5016.162 &Ti I  &...  &...  &...  &...  &...  &...  &5.3  &...  &...  &...  &...       \\
5020.028 &Ti I  &...  &...  &...  &...  &...  &...  &5.0  &...  &...  &...  &...       \\
5022.871 &Ti I  &...  &...  &...  &...  &...  &...  &9.7  &...  &...  &...  &...       \\
5035.907 &Ti I  &...  &5.5  &...  &...  &...  &...  &9.2  &5.3  &...  &...  &...       \\
5036.468 &Ti I  &...  &...  &...  &5.0  &...  &...  &7.0  &...  &...  &...  &...       \\
5039.959 &Ti I  &...  &...  &...  &...  &...  &...  &7.9  &...  &...  &...  &...       \\
5064.654 &Ti I  &...  &...  &...  &...  &...  &...  &8.0  &...  &...  &...  &...       \\
\\		 		 					         	  
4501.272 &Ti II &...  &...  &13.6 &...  &34.0 &24.4 &...  &38.1 &51.5 &29.9 &13.7       \\
4563.761 &Ti II &...  &...  &11.7 &...  &26.5 &19.6 &...  &30.8 &42.7 &24.6 &13.0       \\
4571.968 &Ti II &...  &...  &15.3 &...  &33.3 &24.8 &...  &36.1 &50.6 &30.7 &13.1       \\
4589.958 &Ti II &9.7  &11.1 &...  &9.7  &6.1  &4.8  &27.8 &6.2  &12.9 &4.5  &...       \\
4657.203 &Ti II &...  &...  &...  &...  &...  &...  &9.7  &...  &...  &...  &...       \\
4779.985 &Ti II &...  &...  &...  &...  &...  &...  &13.1 &...  &5.2  &...  &...       \\
5129.152 &Ti II &9.3  &6.5  &...  &...  &...  &...  &19.0 &...  &7.4  &...  &...       \\
5154.070 &Ti II &...  &5.4  &...  &6.5  &...  &...  &13.8 &...  &...  &...  &...       \\
5188.680 &Ti II &13.3 &21.0 &...  &16.5 &5.4  &4.5  &...  &13.4 &24.9 &12.2 &...       \\
5226.543 &Ti II &12.3 &12.4 &...  &9.8  &7.3  &6.1  &34.9 &7.5  &12.8 &...  &...       \\
5336.781 &Ti II &7.6  &5.2  &...  &6.3  &...  &...  &16.5 &...  &5.2  &...  &...       \\
5381.018 &Ti II &...  &...  &...  &...  &...  &...  &...  &...  &...  &...  &...       \\

\enddata
\end{deluxetable}

\end{document}